\documentclass[journal]{IEEEtran}
\usepackage{listings}
\usepackage{cancel}
\usepackage{subfigure}
\usepackage{amsmath}
\usepackage{algorithm, algorithmicx, algpseudocode}
\usepackage[noend]{algcompatible}

\input epsf
\usepackage{graphicx}
\usepackage{enumerate}
\usepackage{color}
\usepackage{amssymb}
\usepackage{tcolorbox}
\usepackage{subfigure} 

\usepackage{stfloats}  

\usepackage{amsmath}   
\usepackage{bm}
\usepackage{amsmath}
\usepackage{xcolor}
\usepackage{kbordermatrix}

\newcommand\bs[1]{\bm{\mathrm{#1}}}

\renewcommand\epsilon{\varepsilon}

\newlength\myindent
\algdef{SE}[SUBALG]{Indent}{EndIndent}{}{\algorithmicend\ }%
\algtext*{Indent}
\algtext*{EndIndent}
\usepackage[noadjust]{cite}
\usepackage{float}
\begin{document}

\title{Efficient Computation of High-order Electromagnetic Field Derivatives for Multiple Design Parameters in FDTD}

\author{Kae-An Liu, ~\IEEEmembership{Student~Member,~IEEE}, Costas~D.~Sarris,~\IEEEmembership{Senior~Member,~IEEE}
\thanks{The authors are  with the Edward S. Rogers Sr. Department of Electrical and
Computer Engineering, University of Toronto, Toronto, ON M5S 3G4, Canada
(e-mail: costas.sarris@utoronto.ca). This research has been supported by the Natural Sciences and Engineering Research Council of Canada.} 
}



%


\maketitle


\begin{abstract}
This paper introduces a new computational framework to derive electromagnetic field derivatives with respect to multiple design parameters up to any order with the Finite-Difference Time-Domain (FDTD) technique. Specifically, only \textit{one} FDTD simulation is needed to compute the first-order field derivatives with respect to $N$ parameters, while \textit{two} FDTD simulations are needed to compute the field derivatives with respect to one parameter up to any order. The field derivatives with respect to $N$ parameters up to any order are computed with $(N+1)$ FDTD runs. In addition to its efficiency, this framework is based on a subtractive cancellation error free approach, providing guaranteed accuracy toward the computation of field derivatives up to any order. With high-order field derivatives available, sensitivity analysis, parametric modelling and uncertainty quantification can be accurately performed.
\end{abstract}

%
\IEEEpeerreviewmaketitle

\section{Introduction}

As the design complexity of microwave circuits and systems increases, the sensitivity analysis with respect to a large number of parameters becomes an essential part in the design process \cite{bandlersevioraims70,pce1,intro-sensitvity-monaco}. Therefore, numerical techniques are in need to find not only the field solution to a given problem, but also its sensitivity under fabrication tolerances and other statistical uncertainties in geometric and material parameters \cite{saltelli-sensitivitybook, jjmartins}.

Notably, sensitivity of any field-based output function of interest can be derived if the corresponding derivatives of the electromagnetic field components are available. For example, the derivatives of the scattering parameters with respect to a design parameter $\xi$, written as $\partial S_{i,j} / \partial \xi$, can be derived by appyling the chain rule to the derivatives of the incident, reflected and transmitted fields at the ports of a microwave structure \cite{costas-csd}.

The standard approach to compute field derivatives is the central finite difference (CFD) method with second-order accuracy. For example, the electric field derivatives of the form $\partial E^{x,y,z} / \partial \xi$ over nominal value $\xi_0$ is approximated as follows:
\begin{equation}
\frac{\partial E^{x,y,z}}{\partial \xi}(\xi_0) = \frac{E^{x,y,z}(\xi_0+h) - E^{x,y,z}(\xi_0-h)}{2h} + \mathcal{O}(h^2) 
\label{CFD}.
\end{equation}
However, this CFD approximation method is subject to subtractive cancellation errors when a small step size $h$ is used \cite{CFD-errorfloor}. In techniques such as FDTD, the geometric sensitivity analysis is often performed by assigning $h$ to the size of the cells corresponding to the geometry, and the step size dilemma is encountered in choosing a small $h$ to minimize the truncation error in (\ref{CFD}) and the dispersion error in mesh distortion, while avoiding subtractive cancellation errors. Moreover, consider a case where the field derivatives with respect to an $N$-dimensional vector of parameters $\bm\xi = \left[ \xi_1, \xi_2, \dots, \xi_N \right]^T$ are of interest. Then, CFD-based methods require at least $2N$ simulations to evaluate these derivatives. Hence, the computation overhead of CFD methods becomes significant when a large number of parameters is considered.

Alternatively, adjoint variable methods (AVM) have been combined with FDTD in \cite{bakr-avm-grid} \cite{bakr-avm}. The AVM-FDTD method computes the gradient of an objective function with respect to  $\bm\xi$ by running one additional FDTD simulation, which is a time-reversed version of the original system \cite{adjoint-network-circuit}\cite{avm-FDTD-korea}. Recently, the AVM-FDTD method was extended to compute second-order derivatives \cite{avm-bakr-2ndorder} by applying FDTD to the wave equation rather than the first-order Maxwell's equations. Indeed, second and higher-order derivatives are very useful in electromagnetic studies such as parametric modelling and optimization of a system \cite{highorder-book,highorder-mem,iceaa2017}. For example, the electric field components can be expanded in a Taylor series with respect to a parameter $\xi$ as: 
\begin{equation}
E^{x,y,z}(\xi) = E^{x,y,z}(\xi_0) + \sum_{n=1}^{\infty} \dfrac{1}{n!}\dfrac{\partial^n E^{x,y,z}(\xi_0)}{\partial \xi^n} (\xi - \xi_0)^n.
\label{taylor}
\end{equation} 
Evidently, parametric design optimization as well as uncertainty quantification analysis can be performed using (\ref{taylor}) \cite{pce1,pce2}.  

To compute field derivatives up to any order accurately, an FDTD-based approach to calculate derivatives without subtractive cancellation errors is proposed in \cite{mcsd-ims2017} using the complex-step derivative (CSD) approximation method. Despite its accuracy, robustness and the compatibility to exiting FDTD codes, the resulting CSD-FDTD method produces a significant computation overhead in memory and execution time when multiple design parameters are considered. On the other hand, \cite{2018ims-fdtd} introduced an alternative approach, by directly differentiating the FDTD update equations. As discussed in \cite{2018ims-fdtd}, the $M$-th order field derivatives can be found recursively from the $(M-1)$-th order field derivative. Hence, the computation overhead scales with the order of derivatives. For example, a total of $M$ simulations is needed to compute the $M$-th order field derivatives with respect to one parameter. 

This paper builds on and significantly extends the work reported in \cite{2018ims-fdtd} and introduces a new framework to compute high-order multi-parametric field derivatives accurately and efficiently. With the proposed approach:
\begin{itemize}
\item First-order field derivatives with respect to $N$ design parameters are found in a single FDTD simulation.
\item Field derivatives up to order $M$ with respect to one parameter require two FDTD simulations.
\item Field derivatives up to order $M$ with respect to $N$ parameters are found in $(N+1)$ FDTD simulations.
\end{itemize}

This is accomplished by mapping the FDTD equations with respect to field derivatives to the original FDTD simulation, driven by ``equivalent sources", placed at Yee cells related to the parameters under consideration. It should be noted that, all the electromagnetic fields in this paper are computed with original FDTD simulations without any mesh distortion.

 
\section{FDTD Update Equations for Field Derivatives}
\label{differential_form}
\subsection{Matrix form of FDTD update equations}
The FDTD update equations are represented in matrix form throughout this paper. In a three dimensional FDTD domain consisting of $N_x \times N_y \times N_z = P$ cells, the electric and magnetic field components in $x,y,z$-directions at all Yee cells are assembled in two vectors $\textbf{E}$ and $\textbf{H}$:
\begin{equation}
\textbf{E} = 
\begin{bmatrix}
\textbf{E}^x \\ \textbf{E}^y \\\textbf{E}^z
\end{bmatrix}
, \textbf{H} = 
\begin{bmatrix}
\textbf{H}^x \\ \textbf{H}^y \\\textbf{H}^z
\end{bmatrix},
\end{equation}
where $\textbf{E}^{x,y,z}$ and $\textbf{H}^{x,y,z}$ are six vectors representing the electric and magnetic fields at all cells. For example:
\begin{equation}
\textbf{E}^x = 
\begin{bmatrix}
E^x_{1,1,1} \\ E^x_{2,1,1} \\ \vdots \\ E^x_{N_x,N_y,N_z}
\end{bmatrix} = \left[ E^x \right]_p \in 	\mathbb{R}^{P \times 1}
\end{equation}
The index notation $p$ of $\textbf{E}^{x,y,z}$ and $\textbf{H}^{x,y,z}$ vectors is a transformation of the triad $(i,j,k)$ used in 3-D FDTD simulations to indicate the location of a Yee cell. The index $p$ is mapped to the triad $(i,j,k)$:
\begin{multline}
\mathbb{P}(i,j,k)  = (k-1)N_xN_y + (j-1)N_x + i, 
\\\forall i = 1,2,..,N_x, j= 1,2,..,N_y, k= 1,2,..., N_z 
\\  = \lbrace p \mid p =1,...,P \rbrace,
  \label{index_trans}
\end{multline}
where $\mathbb{P}$ is the set of all the cell indices in the simulation domain.

In addition to the field vectors, three vectors $\textbf{d}^{x,y,z}$ are defined to represent the reciprocal of the size of each cell. In an FDTD simulation of the geometry under consideration, $\textbf{d}^{x,y,z}$ are vectors of length $M$ with entries $1/\Delta x_p$, $1/\Delta y_p$ and $1/\Delta z_p$ respectively:
\begin{equation}
\textbf{d}^x = 
\begin{bmatrix}
1/\Delta x_1 \\ 1/\Delta x_2 \\ \vdots \\ 1/\Delta x_P
\end{bmatrix},
\textbf{d}^y = 
\begin{bmatrix}
1/\Delta y_1 \\ 1/\Delta y_2 \\ \vdots \\ 1/\Delta y_P
\end{bmatrix},
\textbf{d}^z = 
\begin{bmatrix}
1/\Delta z_1 \\ 1/\Delta z_2 \\ \vdots \\ 1/\Delta z_P
\end{bmatrix}
\in 	\mathbb{R}^{P \times 1}
\end{equation} 

In this paper, it is assumed that the size of a set of cells depends on a specific geometric parameter. That is, the $\Delta x,y,z$ of these cells are functions of a parameter.
Consider the layout of a 3-D microstrip filter with three cascaded stubs, shown in Fig. \ref{zoomin}. The lengths of the stubs are set to nominal values $\bar{\xi_1}, \bar{\xi_2}$ and $\bar{\xi_3}$ at first. If the length of the first stub is a variable $\xi_1$ and deviated by $\delta \xi_1$, then $\xi_1 = \bar{\xi_1} + \delta \xi_1$. Keeping $\delta \xi_1 \rightarrow 0$, only the size of the cells coloured red in the inset of Fig. \ref{zoomin} would be affected by this perturbation. Hence, their new length $\tilde{\Delta y_p}$ in the $y$-direction is written as a function of $\xi_1$:
\begin{equation}
\tilde{ \Delta y_p } (\xi_1)  = \Delta y_p + \delta \xi_1 = \Delta y_p + (\xi_1 - \bar{\xi_1}).
\end{equation}
These cells correspond to the end of the first stub and form a subset region $P_{\xi_1}$ of the FDTD simulation domain, i.e., $P_{\xi_1} \subset \mathbb{P}$. In this example, the vector $\textbf{d}^{y}$ is modified and written as:
\begin{equation}
\textbf{d}^y= \left[ d^y \right]_p =
\begin{cases}
1/\tilde{\Delta y_p}= 1/(\Delta y_p+ \delta \xi_1), p \in P_{\xi_1}\\
1/\Delta y_p, \text{otherwise.}
\end{cases}
\end{equation} 
In this particular example, the $x$ and $y$-component of the size of the cells of subset $P_{\xi_1}$ are independent of $\xi_1$. Hence, $\textbf{d}^{x,z}$ are not modified.
\subsection{FDTD update equations for first-order field derivatives}
\begin{figure}[]
	\centering
    \includegraphics[width=8.2cm]{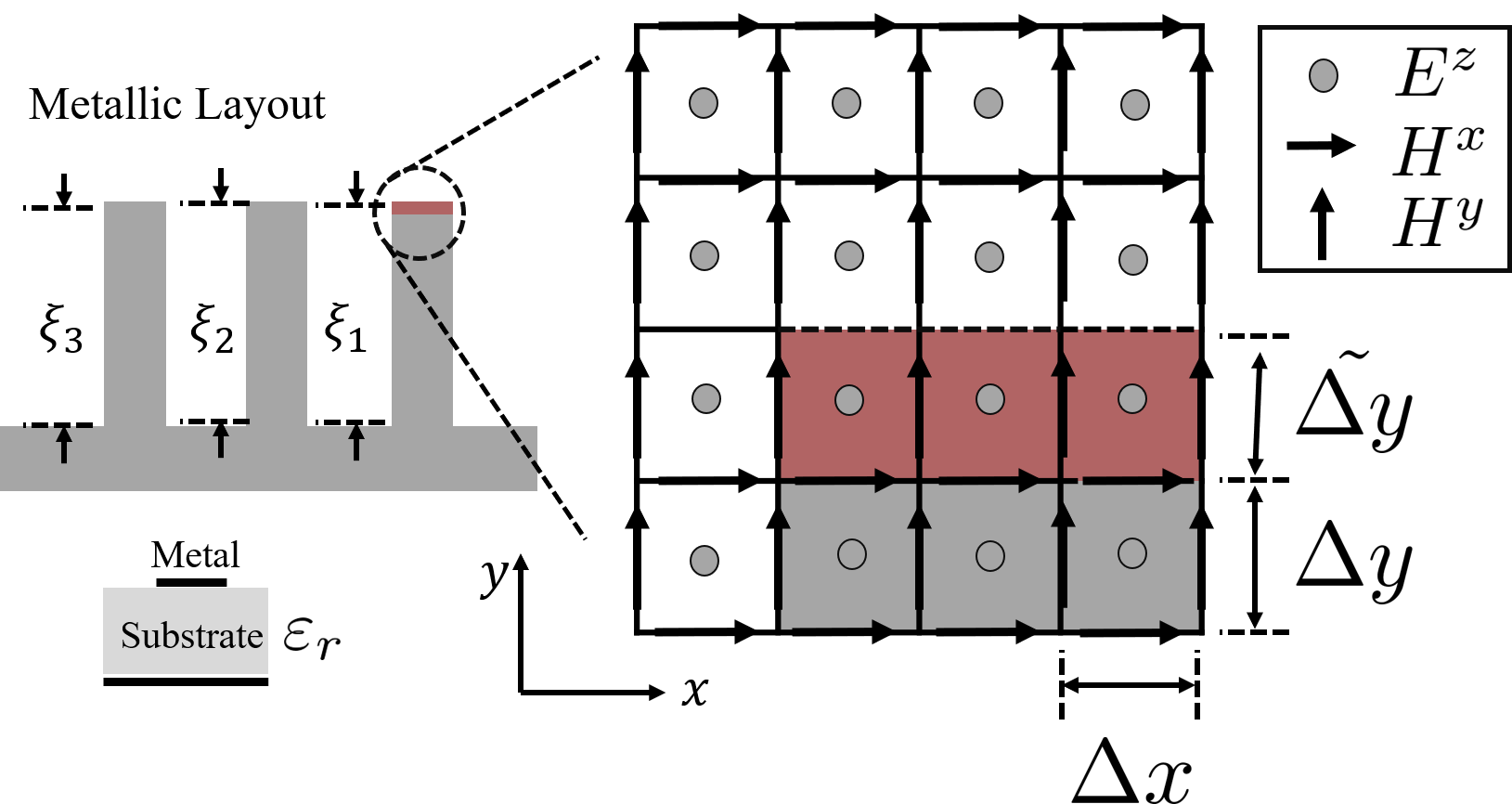}
	\caption{The layout of a microstrip filter with three cascaded stubs. The perturbation of the length of the first stub $\xi_1$ is mapped to the $y$-component of the size of the cells coloured red.}
	\label{zoomin}
\end{figure} 
Consider the 3-D microstrip filter shown in Fig.\ref{zoomin}. The standard FDTD update equations in matrix form are \cite{Gedneybook}:
\begin{equation}
\begin{cases}
\textbf{E}^{n+1} &= \textbf{E}^{n} + \textbf{A}_e \textbf{H}^{n+\frac{1}{2}} \\
\textbf{H}^{n+\frac{1}{2}} &= \textbf{H}^{n-\frac{1}{2}} + \textbf{A}_h\textbf{E}^{n}
\end{cases},
\label{fdtd_eq}
\end{equation}
where $\textbf{A}_e$ and $\textbf{A}_h$ are the update matrices for $\textbf{E}$ and $\textbf{H}$ respectively:
\begin{align}
\textbf{A}_e &= \Delta t \textbf{M}^{-1}_{\epsilon}\left( \textbf{C}e \circ \textbf{D} \right) \\
\textbf{A}_h &= \Delta t \textbf{M}^{-1}_{\mu}\left( \textbf{C}h \circ \textbf{D} \right).
\label{update_mx}
\end{align}
The $\circ$ operator denotes the element-wise multiplication of matrices; $\Delta t$ is the time-step used in the FDTD simulation; $\textbf{M}_{\mu}$ and $\textbf{M}_{\epsilon}$ are the diagonal matrices representing material properties. $\textbf{C}e$ and $\textbf{C}h$ are two sparse matrices defining the discrete curl operators in matrix form with non-zero entities of either $+1$ or $1$. $\textbf{D}$ is a matrix referring to the size of cells and defined as:
\begin{equation}
\textbf{D} \equiv 
\begin{bmatrix}
\textbf{0} & \textbf{D}^{z} &  \textbf{D}^{y}\ \\ 
\textbf{D}^z & \textbf{0} &  \textbf{D}^{x} \\
\textbf{D}^y & \textbf{D}^{x} &  \textbf{0}
\end{bmatrix}
\end{equation}
The block matrices $\textbf{D}^{x,y,z}$ in $\textbf{D}$ are diagonal matrices assembled from vectors $\textbf{d}^{x,y,z}$. Following the derivation in the previous subsection,    $P_{\xi_1}$ is the subset including all the cells with the size depends on small change in $\xi_1$. $\textbf{D}^{y}$ is written as:
\begin{equation}
\textbf{D}^{y} = \text{diag}( \textbf{d}^y) = \left[ D^y \right]_{ij} = \begin{cases} 
    1/\tilde{\Delta y_i}, & \text{if $i=j \in P_{\xi_1}$}\\
    1/\Delta y_i & \text{if $i=j \notin P_{\xi_1}$}\\
    0, & \text{otherwise}.
  \end{cases}
\end{equation}
Again, since the small change or perturbation in $\xi_1$ affects the cell size in the $y$-direction only, $\textbf{D}^x=\text{diag}(\textbf{d}^x)$ and $\textbf{D}^z=\text{diag}(\textbf{d}^z)$ are two diagonal matrices with non-zero, constant entities $1/\Delta x_p$ and $1/\Delta z_p$ respectively.

To find the field derivatives with respect to $\xi_1$, differentiation is performed directly to (\ref{fdtd_eq}), yielding:
\begin{align}
&\begin{cases}
\dfrac{\partial \textbf{E}^{n+1}}{\partial \xi_1} = 
\dfrac{\partial \textbf{E}^{n}}{\partial \xi_1} + 
\textbf{A}_e \dfrac{\partial \textbf{H}^{n+\frac{1}{2}}}{\partial \xi_1} + 
\bm{\mathcal{J}}
\\[.2cm]
\dfrac{\partial \textbf{H}^{n+\frac{1}{2}}}{\partial \xi_1} = 
\dfrac{\partial \textbf{H}^{n-\frac{1}{2}}}{\partial \xi_1} + 
\textbf{A}_h \dfrac{\partial \textbf{E}^{n}}{\partial \xi_1} + 
\bm{\mathcal{K}}
\end{cases}
\label{diff_fdtd_eq}
\\[.2cm]
&\bm{\mathcal{J}}= \dfrac{\partial \textbf{A}_e}{\partial \xi_1} \textbf{H}^{n+\frac{1}{2}}
, \bm{\mathcal{K}} = \dfrac{\partial \textbf{A}_h}{\partial \xi_1} \textbf{E}^{n}
\label{eq_src}
\end{align}

Eq. (\ref{diff_fdtd_eq}) is the FDTD update equation with respect to first-order field derivatives. This differentiated form of FDTD update equations builds the connection between the computation of fields $\left\lbrace \textbf{E},\textbf{H} \right\rbrace$ and their derivatives. Indeed, $\left\lbrace \textbf{E},\textbf{H} \right\rbrace$ are discrete space-time fields, but also functions of $\xi_1$, assumed to be smooth for their derivatives to be well defined \cite{Krumpholz}. Hence:
\begin{equation}
\begin{cases}
E(x,y,z,t; \xi_1) = \displaystyle{\sum_{i,j,k,n}} E^n_{i,j,k}(\xi_1)h_i(x)h_j(y)h_k(z)h_n(t)\\
H(x,y,z,t; \xi_1) = \displaystyle{\sum_{i,j,k,n}} H^n_{i,j,k}(\xi_1)h_i(x)h_j(y)h_k(z)h_n(t),
\end{cases}
\end{equation}
and their derivatives are derived as:
\begin{equation}
\begin{cases}
\dfrac{\partial E(x,y,z,t; \xi_1)}{\partial \xi_1}=
\displaystyle{\sum_{i,j,k,n}} \dfrac{\partial E^n_{i,j,k}(\xi_1)}{\partial \xi_1}h_i(x)h_j(y)h_k(z)h_n(t)\\
\dfrac{\partial H(x,y,z,t; \xi_1)}{\partial \xi_1}=
\displaystyle{\sum_{i,j,k,n}} \dfrac{\partial H^n_{i,j,k}(\xi_1)}{\partial \xi_1}h_i(x)h_j(y)h_k(z)h_n(t),
\end{cases}
\end{equation}
where $h_{i,j,k}$ are pulse functions. For example:
\begin{align}
h_i(x)=h(\dfrac{x}{\Delta x_i}-i), \text{and} \\
h(\tau)=
\begin{cases}
1, \dfrac{-1}{2}\leq \tau\leq \dfrac{1}{2} \\
0, \text{otherwise}.
\end{cases}
\end{align}

Further examination of (\ref{diff_fdtd_eq}) reveals that the electric and magnetic field derivatives are updated in time, using the original update matrices $\textbf{A}_{e,h}$. In addition, the fields derived from previous time steps appear as ``equivalent sources" $\bm{\mathcal{J}}$ and $\bm{\mathcal{K}}$ in the update equations for field derivatives.
\subsection{Equivalent sources for field derivatives}
As stated in (\ref{eq_src}), the equivalent sources $\bm{\mathcal{J}}$ and $\bm{\mathcal{K}}$ for the first-order field derivatives with respect to $\xi_1$ are products of $\partial \textbf{A}_{e,h}/\partial \xi_1$ and fields $\textbf{E}$ and $\textbf{H}$. The derivatives of update matrices $\textbf{A}_{e,h}$ are further expressed as:
\begin{equation}
\dfrac{\partial \textbf{A}_{e,h}}{\partial \xi_1} = \Delta t \textbf{M}^{-1}_{\epsilon, \mu}(\textbf{C}_{e,h} \circ \dfrac{\partial \textbf{D}}{\partial \xi_1}).
\end{equation}
The derivative of the matrix $\textbf{D}$ with respect to $\xi_1$ is a sparse matrix:
\begin{equation}
\begin{split}
\dfrac{\partial \textbf{D}}{\partial \xi_1} &= 
\begin{bmatrix}
\textbf{0} & \partial \textbf{D}^{z} / \partial \xi_1 &  \partial \textbf{D}^{y} / \partial \xi_1\ \\ 
\partial \textbf{D}^{z} / \partial \xi_1 & \textbf{0} &  \partial \textbf{D}^{x} / \partial \xi_1 \\
\partial \textbf{D}^{y} / \partial \xi_1 & \partial \textbf{D}^{x} / \partial \xi_1 &  \textbf{0}
\end{bmatrix}
\\[.2cm]&=
\begin{bmatrix}
\textbf{0} & \textbf{0} &  \partial \textbf{D}^{y} / \partial \xi_1 \\ 
\textbf{0} & \textbf{0} &  \textbf{0} \\
\partial \textbf{D}^{y} / \partial \xi_1 & \textbf{0} &  \textbf{0}
\end{bmatrix}.
\end{split}
\end{equation}
Since the perturbation in $\xi_1$ affects the size of the cells in the $y$-direction only and all entries in $\textbf{D}^{x,z}$ are independent of this perturbation. Hence, $\partial \textbf{D}^{x}/ \partial \xi_1 = \textbf{0}$ and $\partial \textbf{D}^{z}/ \partial \xi_1 = \textbf{0}$. 
On the other hand, the entities in $\textbf{D}^y$ are either constant $0$, $1/\Delta y_p$ and variable $1/\tilde{\Delta y_p}$. Notably, only the derivative of $1/\tilde{\Delta y_p}$ with respect to $\xi_1$ is not zero: 
\begin{multline}
\dfrac{\partial  \left( \dfrac{1}{\tilde{\Delta y_p}}\right)}{\partial \xi_1}=
\dfrac{\partial  \left( \dfrac{1}{\Delta y_p + \xi_1 - \bar{\xi_1}}\right)}{\partial \xi_1}
\\=\dfrac{-1}{\left( \Delta y_p + \xi_1 - \bar{\xi_1} \right)^2}
\approx 
\dfrac{-1}{{\Delta y_p}^2}.
\label{core}
\end{multline}
The following equation summarizes the derivative of $\textbf{D}^y$ with respect to $\xi_1$:
\begin{equation}
\dfrac {\partial \textbf{D}^{y}}{\partial \xi_1} =
\left[ \dfrac{\partial D^y}{\partial \xi_1} \right]_{ij} = 
\begin{cases}
    1/\Delta y_i^2, & \text{if $i=j \in P_{\xi_1}$},\\
    0, & \text{otherwise}.
  \end{cases}
\label{core2}
\end{equation}

The derivations in (\ref{core}) and (\ref{core2}) indicate that the derivatives of update matrices $\textbf{A}_{e,h}$ are analytically available. Therefore, the equivalent source exciting the first-order field derivatives can be found with \textbf{one} standard FDTD simulation, exploiting the fields computed by this FDTD simulation and the analytical expression of the derivatives of $\textbf{A}_{e,h}$.

In this 3-D example, the vectors $\bm{\mathcal{J}}$ and $\bm{\mathcal{K}}$ are further written as:
\begin{equation}
\bm{\mathcal{J}} = \begin{bmatrix}
\bm{\mathcal{J}}^x \\ \bm{\mathcal{J}}^y \\\bm{\mathcal{J}}^z
\end{bmatrix}, 
\bm{\mathcal{K}} = \begin{bmatrix}
\bm{\mathcal{K}}^x \\ \bm{\mathcal{K}}^y \\\bm{\mathcal{K}}^z
\end{bmatrix},
\end{equation}
where $\bm{\mathcal{J}}^{x,y,z}$ and $\bm{\mathcal{K}}^{x,y,z}$ are the vectors of equivalent sources for first-order field derivatives in $x,y,z$-directions respectively.

\begin{figure}[]
	\centering
    \includegraphics[width=8cm]{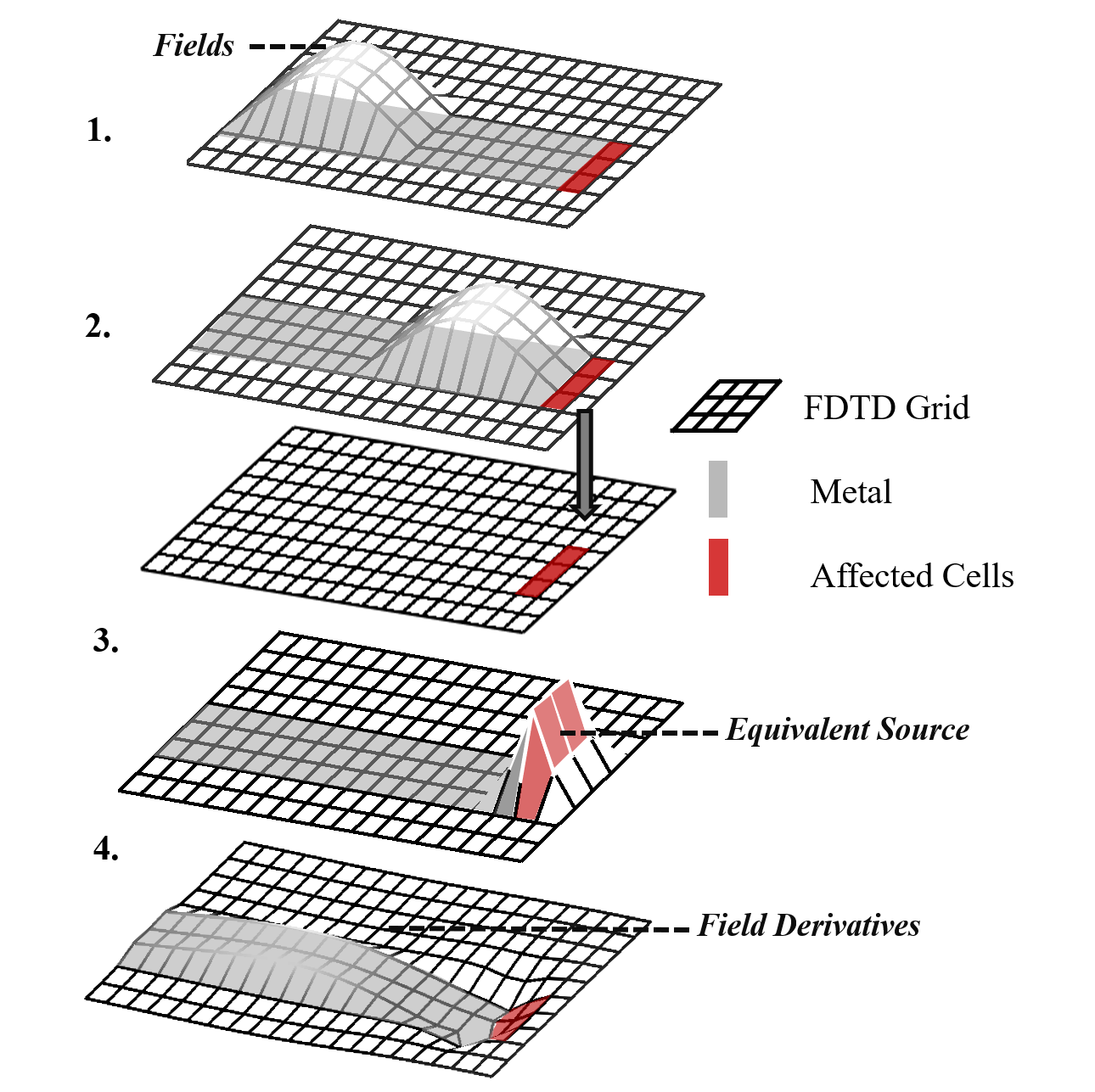}
	\caption{The equivalent sources for field derivatives are computed using the fields solved in an original FDTD simulation and the derivatives of update matrices. The derivatives of update matrices have non-zero entities, only if the entities are affected by the perturbation of a design parameter. The field derivatives are then excited by the equivalent source.}
	\label{sparese_matrix_exc}
\end{figure}

Fig. \ref{sparese_matrix_exc} illustrates the full-wave computation of the equivalent sources in FDTD. When the fields reach the $P_{\xi_1}$ cells coloured red in the grid, the field derivatives are excited. Notably, the derivatives of $\textbf{A}_{e,h}$ and the fields probed at these cells are used to compute the equivalent source for the field derivatives, which further propagate throughout the grid.

\section{Multi-parametric Sensitivity and Equivalent Sources}
\label{Multi-parametric}
\subsection{Jacobian matrix of field derivatives}

\begin{figure}[]
	\centering
    \includegraphics[width=8cm]{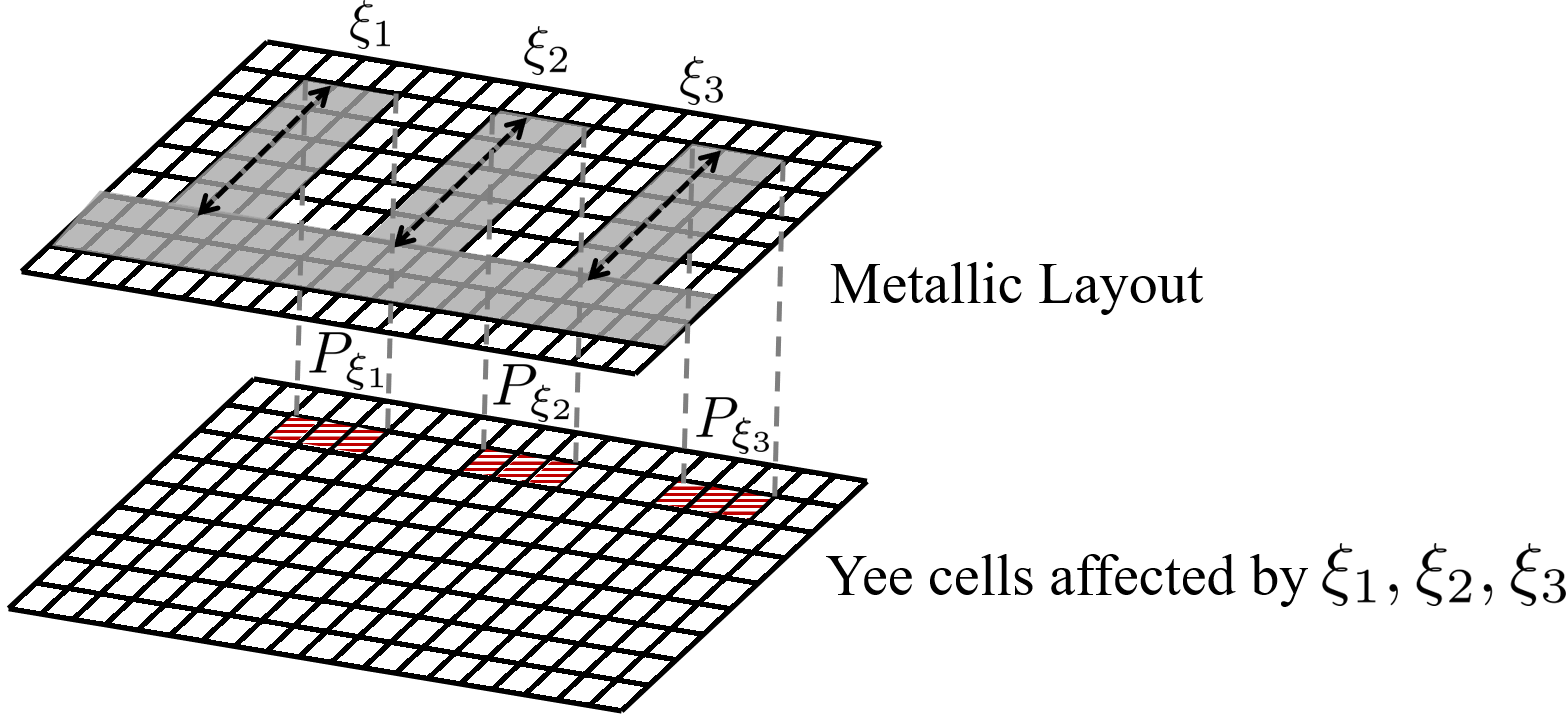}
	\caption{The size of the cells filled with slash stripes depends on the perturbation of stub lengths $\xi_1, \xi_2, \xi_3$. }
	\label{sparese_matrix}
\end{figure} 


Section \ref{differential_form} proposed a new approach to find the relation of fields and field derivatives in FDTD. 
In addition, the equivalent source for field derivatives with respect to parameter $\xi$ is located only at the cells of $P_\xi$.

This section extends this approach to the multiparametric case. Consider the derivative of the electric field in the $z$-direction sampled at a cell of index $p_a$ with respect to an $N$-dimesional vector of design parameters $\bs{\xi} = \left[ \xi_1, \xi_2, \cdots \xi_N  \right]^T$:
\begin{equation}
\nabla_{\bs{\xi}}{{E}^z_{p_a}} \equiv
 \begin{bmatrix}
 \dfrac{\partial {E^{z}_{p_a}}}{\partial \xi_1} & \dfrac{\partial {E^{z}_{p_a}}}{\partial \xi_2} & \dots  &  \dfrac{\partial {E^{z}_{p_a}}}{\partial \xi_N} 
\end{bmatrix},
\label{jacobian_matrix}
 \end{equation}
which is a $1\times N$ Jacobian matrix. 

Assume perturbations $\delta_1, \delta_2, ..., \delta_N$ of $\xi_1, \xi_2..., \xi_N$ affect the $x$-component of the cell-size for cells $p_1, p_2, ..., p_N$ respectively. Hence, the vector $\textbf{d}^x$ in this 3-D FDTD simulation is written as:
\begin{equation}
\setlength\arraycolsep{2pt}
\begin{split}
&\textbf{d}^x= \left[ d^x \right]_p = 
\\
&
\kbordermatrix{
& 1 & 2 & p_1 & \dots & p_2 & \dots & p_N & \dots & P \\
& \frac{1}{\Delta x_1} & \frac{1}{\Delta x_2} &\frac{1}{\Delta x_{p_1}+\delta_1} &\dots  &\frac{1}{\Delta x_{p_2}+\delta_2} & \dots  &\frac{1}{\Delta x_{p_N}+\delta_N} & \dots& \frac{1}{\Delta x_P} }
\label{dff_dz}
\end{split}
\end{equation}
In addition, an operator $\mathfrak{D}$ is defined to differentiate the entries of FDTD update matrices that are related to the parameters $\bm{\xi}$. In this example  $\mathfrak{D}_{\bm{\xi}}$ is:
\[
  \mathfrak{D}_{\bm{\xi}}= \kbordermatrix{
    & 1 & 2 & p_1 & \dots & p_2 & \dots & p_N & \dots & P \\
    1 & 0 & 0 & 0 & 0 & 0 & 0 & 0 & 0 & 0\\
    2 & 0 & 0 & 0 & 0 & 0 & 0 & 0 & 0 & 0\\
    p_1 & 0 & 0 & \frac{\partial}{\partial \xi_1} & 0 & 0 & 0 & 0 & 0 & 0 \\
    \vdots & \vdots & \vdots & \vdots & \ddots & \vdots & \ddots & \vdots & \ddots & \vdots\\
    p_2 & 0 & 0 & 0 & 0 & \frac{\partial}{\partial \xi_2} & 0 & 0 & 0 & 0 \\
    \vdots & \vdots & \vdots & \vdots & \ddots & \vdots & \ddots & \vdots & \ddots & \vdots\\
    p_N & 0 & 0 & 0 & 0 & 0 & 0 & \frac{\partial}{\partial \xi_N} & 0 & 0 & \\
    \vdots & \vdots & \vdots & \vdots & \ddots & \vdots & \ddots & \vdots & \ddots & \vdots\\
    P & 0 & 0 & 0 &  0 & 0 & 0 & 0 & 0 & 0  
  }
\]
Therefore, the derivative of $\textbf{D}^x = \text{diag}(\textbf{d}^x)$ with respect to $\bm{\xi}$ is found as:
\begin{equation}
\mathfrak{D}_{\bm{\xi}}\textbf{D}^x = \mathfrak{D}_{\bm{\xi}} \left[ D^x \right]_{ij} =
\begin{cases}
-1/{\Delta x_i}^2, i = j = p_1,p_2,...,p_N
\\
0, \text{otherwise}.
\end{cases}
\label{D_DX}
\end{equation}
Then, a compact form of the equivalent sources for field derivatives with respect to $\bm{\xi}$ is derived. Consider, for example, the equivalent sources for electric field derivatives $\bm{\mathcal{J}}$:
\begin{equation}
\begin{split}
\bm{\mathcal{J}} &= \Delta t\textbf{M}^{-1}_{\epsilon}\left( \textbf{C}e \circ 
\begin{bmatrix}
\textbf{0} &\textbf{0} &\textbf{0} \\
\textbf{0} &\textbf{0} &\mathfrak{D}_{\bm{\xi}}\textbf{D}^x\\
\textbf{0} &\mathfrak{D}_{\bm{\xi}}\textbf{D}^x &\textbf{0}
\end{bmatrix}
\right)\textbf{H}
\\[.3cm]
&=\begin{bmatrix}
\bm{\mathcal{J}}^x \bm{\mathcal{J}}^y \bm{\mathcal{J}}^z
\end{bmatrix}^T
\end{split},
\label{J_DX}
\end{equation}
where the vector $\bm{\mathcal{J}}^z$ contains equivalent sources for the $\textbf{E}^z$ derivatives. Substituting (\ref{D_DX}) into (\ref{J_DX})  , the vector $\bm{\mathcal{J}}^z$ is defined in the time domain as:
\begin{multline}
\bm{\mathcal{J}}^z (t) = \left[ \mathcal{J}^z(t) \right]_p \\=
\begin{cases}
\dfrac{- \Delta t}{\epsilon_0 \epsilon_r {\Delta x_p}^2}
\left( H^y_{p}(t) - H^y_{p-1}(t) \right), p = p_1, p_2, ..., p_N
\\
0, \text{otherwise}.
\end{cases}
\label{j_first_order}
\end{multline}
The non-zero entries in $\bm{\mathcal{J}}^z$: $\mathcal{J}^z_{p_1}, \mathcal{J}^z_{p_2}, ...,\mathcal{J}^z_{p_N}$ are the equivalent sources for field derivatives with respect to $\xi_1,\xi_2,...,\xi_N$ respectively. Eq. (\ref{j_first_order}) indicates that the equivalent sources for field derivatives with respect to multiple parameters can be computed simultaneously with the original FDTD simulation.

It was shown before that the derivatives of $\text{E}^z$ with respect to $\xi_1$ are excited by $\mathcal{J}^z_{p_1}$ located at cell $p_1$; therefore, the field derivative at an observation port $p_a$, such as the input or output port of a system, can be found by using the following relation in the frequency domain:
\begin{equation}
\dfrac{\partial \hat{E^{z}_{p_a}}}{\partial \xi_1}(f) = \mathcal{H}_{p_ap_1}(f) \cdot
\hat{\mathcal{J}^{z,1}_{p_1}} (f),
\label{s21_ts}
\end{equation}
where $\mathcal{H}_{p_ap_1}$ is the transfer function from cell $p_1$ to port $p_a$ and superscript $\hat{}$ denotes a Fourier transform. $\hat{E}^{z}$ and $\hat{\mathcal{J}}^z$ are the $z$-component of the Fourier-transformed electric fields and of the equivalent source for electric field derivatives. In general, the transfer functions between port $p_a$ and the cells at $p_1,p_2,\dots,p_N$ are found by exciting the grid with a Gaussian source placed at $p_a$ and recording the fields at $p_a$ and $p_n$ over time :
\begin{equation}
\mathcal{H}_{p_ap_n}(f)=  \dfrac {\mathfrak{F}\left[ E^{z}_{p} (t) \right] }{ \mathfrak{F}\left[ E^{z}_{p_a} (t) \right]}, p_n = p_1,p_2,...,p_N,
\label{transfer_fnt}
\end{equation}
where $\mathfrak{F}$ denotes Fourier transform. With these transfer functions and equivalent sources available, $\nabla_{\bs{\xi}}{\hat{E}^z_{p_a}}$ is written as:
\begin{equation}
\nabla_{\bs{\xi}}{\hat{E}^z} = \bm{\mathcal{H}}\bm{\mathcal{G}}^T
\label{multi-full}
\end{equation}
\begin{equation}
\bm{\mathcal{H}} \equiv \begin{bmatrix} 
\mathcal{H}_{p_ap_1} & \mathcal{H}_{p_ap_2} &\dots & \mathcal{H}_{p_ap_N} \end{bmatrix}
\end{equation}
\begin{equation}
\bm{\mathcal{G}} \equiv \begin{bmatrix} 
\hat{\mathcal{J}}^{z}_{p_1} & \hat{\mathcal{J}}^{z}_{p_2} &\dots & \hat{\mathcal{J}}^{z}_{p_N} \end{bmatrix}
\end{equation}
In (\ref{multi-full}), the following relation holds for each element of the matrix $\nabla_{\bs{\xi}}{\hat{E}^z} $:
\begin{equation}
 \dfrac{\partial \hat{E^{z}_{p_a}}}{\partial \xi_n}(f) = 
S_{p_ap_n}(f) \cdot
\hat{\mathcal{J}}^{z,1}_{p_n} (f), n = 1,2,...,N
\label{multi-relation}
\end{equation}
Eq. (\ref{multi-full}) indicates that the Jacobian matrix of the field at a cell of interest, which is $p_a$ here, can be computed by running one FDTD simulation and exploiting the transfer functions and equivalent sources directly.  

In general design problems, where the perturbation of a parameter $\xi_n$ is relevant to multiple cells, these affected cells form a subset $P_{\xi_n} \subset P$. 
For example, Fig. \ref{sparese_matrix} illustrates three subsets of the cells affected by the perturbation in the length of the stubs of the example shown in Fig. \ref{zoomin}.  
Therefore, in such cases, the source is distributed over a surface or a volume and (\ref{multi-relation}) is modified as follows:
\begin{equation}
\dfrac{\partial \hat{E^{z}_{p_a}}}{\partial \xi_n}(f) =  \displaystyle{\sum_{p_i \in P_{\xi_n}}} \mathcal{H}_{p_ap_i}(f)\cdot \hat{\mathcal{J}}^{z}_{p_i} (f), n = 1,2,...,N.
\label{eq_relation1}
\end{equation}

\subsection{Implementation}
To numerically implement the proposed method, the FDTD update equations and equivalent sources derived in (\ref{diff_fdtd_eq}) are discretized. For example, the electric fields in the $z$-direction are updated as:
\begin{multline}
E_{i,j,k}^{z,n+1} =  E_{i,j,k}^{z,n}  + {A_e^x}_{i,j,k}   \left(
H_{i+\frac{1}{2},j,k}^{y,n+\frac{1}{2}}- 
H_{i-\frac{1}{2},j,k}^{y,n+\frac{1}{2}} \right)
\\[.1cm]
- {A_e^y}_{i,j,k}\left(
H_{i,j+\frac{1}{2},k}^{x,n+\frac{1}{2}}-
H_{i,j-\frac{1}{2},k}^{x,n+\frac{1}{2}} \right),
\label{fdtd_eq}
\end{multline} 
where ${A_e^x}$ and ${A_e^y}$ are the update coefficients:
\begin{equation}
\begin{split}
{A_e^x}(i,j,k) = \dfrac{\Delta t}{M_{\epsilon}(i,j,k)\cdot d^x(i,j,k)}\\
{A_e^y}(i,j,k) = \dfrac{\Delta t}{M_{\epsilon}(i,j,k)\cdot d^y(i,j,k)}\\
{A_e^z}(i,j,k) = \dfrac{\Delta t}{M_{\epsilon}(i,j,k)\cdot d^z(i,j,k)}.
\end{split}
\end{equation}
$M_{\epsilon}$, $d^{x}$, $d^{y}$, $d^{z}$ are three-dimensional matrices represents the permittivity and cell size. In addition, the equivalent sources for the derivatives of the $z$-component of the electric field, with respect to $\xi_n$, are updated as:
\begin{multline}
\mathcal{J}^{z,n}({i,j,k}) = \dfrac{\partial {A_e^x}(i,j,k)}{\partial \xi_n} \left(
H_{i+\frac{1}{2},j,k}^{y,n+\frac{1}{2}}- 
H_{i-\frac{1}{2},j,k}^{y,n+\frac{1}{2}} \right)
\\ - \dfrac{\partial {A_e^y}(i,j,k)}{\partial \xi_n}\left(
H_{i,j+\frac{1}{2},k}^{x,n+\frac{1}{2}}-
H_{i,j-\frac{1}{2},k}^{x,n+\frac{1}{2}} \right).
\label{implement-eq}
\end{multline}
If the cell located at $(i,j,k) \in P_{\xi_n}$, the derivatives of the corresponding update coefficients are as follows:
\begin{equation}
\dfrac{\partial {A_e^x} (i,j,k)}{\partial \xi_n} = 
\begin{cases}
\dfrac{-\Delta t}{M_{\epsilon}(i,j,k)\cdot [{d^x}(i,j,k)]^2}, (i,j,k) \in P_{\xi}\\
0, \text{ otherwise}.
\end{cases}
\end{equation}
Meanwhile, the derivatives of update coefficients in $y$- and $z$-directions are zero throughout the simulation domain:
\begin{equation}
\\[.2cm]
\dfrac{\partial {A_e^y}(i,j,k)}{\partial \xi_n} = \dfrac{\partial {A_e^z}(i,j,k)}{\partial \xi_n}= 0,    \forall (i,j,k)
\end{equation}
The discretized update equations for magnetic fields and equivalent sources $\mathcal{K}$ share the same structure as of (\ref{fdtd_eq}) and (\ref{implement-eq}) with $M_{\varepsilon}$ replaced by $M_{\mu}$. 

The steps for this method are then presented in the Algorithm 1. This algorithm includes three parts: (1) Mapping perturbations of design parameters to FDTD update coefficients, (2) FDTD simulation and (3) post-processing. The computation overhead for the first and third parts is negligible compared to an FDTD run.

\begin{algorithm}
\label{firstalgo}
 \caption{Compute Jacobian Matrix of Field Derivatives with Equivalent Sources in FDTD}
 \begin{algorithmic}[1]
  \Statex $P_1,P_2,...,P_N$: subsets of the cells affected by parameters $\xi_1, \xi_1, ...\xi_N$
  \Statex $p_a$: observation port
  \Statex $p$: matrix index mapping triad $(i,j,k)$, $p = 1,2,\dots,P$
\\ \textbf{procedure} \textsc{Perturbation Mapping}
  \Indent
  \FOR {$p = 1$ to $P$}
    \IF {$ p \in P_n, n = 1 :N$}
  \STATE ${A^{x,y,z}_e(p)}' \gets \frac{\partial A^{x,y,z}_e(p)}{\partial \xi_n} $
  \STATE ${A^{x,y,z}_h(p)}' \gets \frac{\partial A^{x,y,z}_h(p)}{\partial \xi_n} $
     \ENDIF
  \ENDFOR
  \EndIndent   
\\ \textbf{procedure} \textsc{FDTD Run}
  \Indent
  \FOR {$t = 1$ to $Tsteps$}
  	\IF {$t<\text{excitation time}$}
  		\STATE $E^z_{p_a} \gets g(t)$ Gaussian excitation
  	\ENDIF
  	\STATE Update $E^{x,y,z}(1:P)$
  	\STATE Update $H^{x,y,z}(1:P)$
  	\STATE Record fields at $p\in P_{\xi_n}, n=1:N$
  	\STATE Compute $\mathcal{J}^{x,y,z}$ and $\mathcal{K}^{x,y,z}$ using (\ref{j_first_order}):
  	\Indent
	  	\STATE $\mathcal{J}^{x,y,z}(1:P) \gets {A^{x,y,z}_e(1:P)}', H^{x,y,z}(1:P) $
	  	\STATE $\mathcal{K}^{x,y,z}(1:P) \gets {A^{x,y,z}_h(1:P)}', E^{x,y,z}(1:P) $
	\EndIndent 
  \ENDFOR   
  \EndIndent
\\ \textbf{procedure} \textsc{Equivalent Source Propagation}
   \Indent
    \FOR {$n=1$ to $N$}
     \STATE Find transfer functions using (\ref{transfer_fnt})
     \Indent
	     \STATE $\mathcal{H}_{p_ap_i} (f)\gets \mathfrak{F}[{E^z_{p_i}(t)}]/\mathfrak{F}[{E^z_{p_a}(t)}], p_i \in P_n$
	 \EndIndent
	 \STATE Propagate $\mathcal{J}^z$ to $p_a$ using (\ref{eq_relation1}) 
	 \Indent   
     \STATE $\frac{\partial \hat{E^z_{P_a}}}{\partial \xi_n}(f) \gets \mathcal{H}_{p_ip_a}(f) \cdot
      \mathfrak{F}[\mathcal{J}^z_{p_i}(t)], p_i \in P_n$
      \EndIndent
    \ENDFOR
    \EndIndent
 \end{algorithmic} 
 \end{algorithm}

\section{High-order field derivatives}
Beyond first-order field derivatives, this section further illustrates how high-order derivatives can be computed by applying the approach proposed in Section \ref{Multi-parametric}. 
By using the equivalent sources for field derivatives and the transfer functions, field derivatives with respect to a parameter $\xi_1$, up to any order can be computed with only \textbf{two} FDTD simulations.
\label{high-order}
\subsection{Equivalent sources for high-order field derivatives}
Differentiating (\ref{diff_fdtd_eq}) with respect to $\xi_1$ once more yields the FDTD update equation with respect to second-order field derivatives:
\begin{align}
&\begin{cases}
\dfrac{\partial^2 \textbf{E}^{n+1}}{\partial \xi_1^2} = 
\dfrac{\partial^2 \textbf{E}^{n}}{\partial \xi_1^2} + 
\textbf{A}_e \dfrac{\partial^2 \textbf{H}^{n+\frac{1}{2}}}{\partial \xi_1^2} + 
\bm{\mathcal{J}}^{(2)}
\\[.2cm]
\dfrac{\partial^2 \textbf{H}^{n+\frac{1}{2}}}{\partial \xi_1^2} = 
\dfrac{\partial^2 \textbf{H}^{n-\frac{1}{2}}}{\partial \xi_1^2} + 
\textbf{A}_h \dfrac{\partial^2 \textbf{E}^{n}}{\partial \xi_1^2} + 
\bm{\mathcal{K}}^{(2)}
\end{cases}
\label{diff_fdtd_2nd}
\end{align}
\begin{multline}
\bm{\mathcal{J}}^{(2)}
 = 2 \dfrac{\partial \textbf{A}_e}{\partial \xi_1} \dfrac{\partial \textbf{H}^{n+\frac{1}{2}}}{\partial \xi_1} +  \dfrac{\partial^2 \textbf{A}_e}{\partial \xi_1^2}\textbf{H}^{n+\frac{1}{2}},
\\
 \bm{\mathcal{K}}^{(2)} = 2 \dfrac{\partial \textbf{A}_h}{\partial \xi_1} \dfrac{\partial \textbf{E}^{n}}{\partial \xi_1} + \dfrac{\partial^2 \textbf{A}_h}{\partial \xi_1^2}\textbf{E}^{n}
 \label{J_2nd_order}
\end{multline}

Apparently, (\ref{diff_fdtd_2nd}) shares the same structure as that of (\ref{diff_fdtd_eq}). The second-order field derivatives are updated through the same update matrices $\textbf{A}_{e,h}$. Equivalent sources $\bm{\mathcal{J}}^{(m)}$ and $\bm{\mathcal{K}}^{(m)}$ are used to excite the second-order electric and magnetic field derivatives respectively. The superscript $m$ denotes the equivalent sources for $m$-th order field derivatives. In (\ref{J_2nd_order}), zero-th and first order field derivatives are used to compute $\bm{\mathcal{J}}^{(2)}$ and $\bm{\mathcal{K}}^{(2)}$. 
Repeatedly differentiating (\ref{diff_fdtd_2nd}), the equivalent source terms that appear in the $m$-th order version of the previous equations, exciting $m$-th order field derivatives with respect to $\xi_1$ are: 
\begin{align}
\bm{\mathcal{J}}^{(m)}  &=  \sum_{q=0}^{m-1} \dfrac{m!}{(m-q)! q!}  
\left(\dfrac{\partial^{m-q} \textbf{A}_e}{{\partial \xi_1}^{m-q}} \right)
\dfrac{\partial^q \textbf{H}}{\partial \xi_1^q}
\\[.2cm]
\bm{\mathcal{K}}^{(m)}  &=  \sum_{q=0}^{m-1} \dfrac{m!}{(m-q)! q!}   
\left(\dfrac{\partial^{m-q} \textbf{A}_h}{{\partial \xi_1}^{m-q}} \right)
\dfrac{\partial^q \textbf{E}}{\partial \xi_1^q}.
\label{JmKm}
\end{align}
In brief, the equivalent sources for $m$-th order field derivatives depend on up to $(m-1)$-th order field derivatives and the $m$-th order derivatives of the update matrices $\textbf{A}_{e,h}$. 

Following the assumption used in (\ref{core2}), the perturbation of $\xi_1$ affects $d^y_p , p\in P_{\xi_1}$ only. Again, $P_{\xi_1}$ is the subset including all the affected cells. Hence, the $m$-th order derivative of the update matrix $\textbf{A}_e$ is derived as:
\begin{equation}
\dfrac{\partial^m \textbf{A}_{e}}{\partial \xi_1^m}=\Delta t \textbf{M}^{-1}_{\epsilon}\left(\textbf{C}e \circ 
\begin{bmatrix}
\bm{0} & \bm{0} & \partial^m \textbf{D}^y/\partial {\xi_1}^m \\
\bm{0} & \bm{0} &\bm{0} \\
\partial^m \textbf{D}^y/\partial {\xi_1}^m & \bm{0} & \bm{0} 
\end{bmatrix}
\right),
\end{equation}
where 
\begin{equation}
\dfrac{\partial^m \textbf{D}^y}{\partial {\xi_1}^m}= \dfrac{\partial^m \left[ \text{diag}(\textbf{d}^y)\right]_{ij}}{\partial {\xi_1}^m}=
\begin{cases}
\dfrac{\partial^m (1/\tilde{\Delta y_i})}{\partial {\xi_1}^m}, i = j \in P_{\xi_1}
\\
0, \text{otherwise}.
\end{cases}
\label{Dy_high}
\end{equation}
Using (\ref{core}), the analytical expression of the non-zero terms in (\ref{Dy_high}) is:
\begin{equation}
\dfrac{\partial^m (1/\tilde{\Delta y_p})}{\partial {\xi_1}^m}
= 
(-1)^m m! (\Delta y_p)^{-(m+1)}.
\end{equation}

Hence, only the terms corresponding to the cells of $P_{\xi_1}$ are non-zero in $\partial^m \textbf{A}_{e,h}/\partial \xi^m$. This indicates that the high-order equivalent sources are localized at the $P_{\xi_1}$ cells only. For example:
\begin{equation}
\bm{\mathcal{J}^{(m)}}=
\begin{bmatrix}
\bm{\mathcal{J}^{x,(m)}} \\\bm{\mathcal{J}^{y,(m)}} \\ \bm{\mathcal{J}^{z,(m)}}
\end{bmatrix}
=
\begin{cases}
\left\lbrace
\mathcal{J}^{x,(m)}_p , \mathcal{J}^{y,(m)}_p , \mathcal{J}^{z,(m)}_p
\right\rbrace, p \in P_{\xi_1}
\\[.3cm]
\left\lbrace
 0 , 0  ,0 
\right\rbrace,\text{otherwise.}
\end{cases}
\end{equation}
Since the equivalent sources $\bm{\mathcal{J}}^{(m)}$ are localized at the $P_{\xi_1}$ cells, they are dependent on $\dfrac{\partial^{m-1} H^{x,y,z}_p}{\partial {\xi_1}^{m-1}}, p \in P_{\xi_1}$, which are the $(m-1)$-th order magnetic field derivatives located at the $P_{\xi_1}$ cells. Therefore, the product terms in (\ref{JmKm}) can be represented as follows:
\begin{multline}
\dfrac{\partial^{m-q} \textbf{A}_e}{\partial \xi_1^{m-q}} \dfrac{\partial^q \textbf{H}}{\partial \xi_1^q} = 
 \dfrac{\partial^{m-q} \textbf{A}_e}{\partial \xi_1^{m-q}} \dfrac{\partial^q \overset{\star}{\textbf{H}}}{\partial \xi_1^q} 
\\ \equiv
\dfrac{\partial^{m-q} \textbf{A}_e}{\partial \xi_1^{m-q}}
\begin{bmatrix}
\begin{bmatrix}
0 \\ 0 \\ \vdots \\ \frac{\partial^q H^x_{p}}{\partial \xi_1^q} \\ \vdots
\end{bmatrix}
&
\begin{bmatrix}
0 \\ 0 \\ \vdots \\ \frac{\partial^q H^y_{p}}{\partial \xi_1^q} \\ \vdots
\end{bmatrix}
&
\begin{bmatrix}
0 \\ 0 \\ \vdots \\ \frac{\partial^q H^z_{p}}{\partial \xi_1^q} \\ \vdots
\end{bmatrix}
\end{bmatrix}^T,
p \in P_{\xi_1},
\label{product_term_in_J}
\end{multline}
where the superscript $\star$ indicates that only the entries corresponding to the cells that would be affected by perturbations of parameters, are non-zero in a matrix. Thus, the computation of the $m$-th order equivalent sources is focused on finding up to $(m-1)$-th order field derivatives at the affected cells only. 

Indeed, the $(m-1)$-th order field derivatives are not available directly after running the original FDTD simulation. However, they are excited by $\bm{\mathcal{J}}^{m-1}$ and $\bm{\mathcal{K}}^{m-1}$. Specifically, using the relation stated in (\ref{eq_relation1}), $ {\partial^{m-1} H^{x,y,z}_{p}}/{\partial \xi^{m-1}} $ is found in the frequency domain as:
\begin{equation}
\begin{bmatrix}
\dfrac{\partial^{m-1} \hat{H}^x_{p}}{\partial \xi_1^{m-1}}\\[.3cm]
\dfrac{\partial^{m-1} \hat{H}^y_{p}}{\partial \xi_1^{m-1}}\\[.3cm]
\dfrac{\partial^{m-1} \hat{H}^z_{p}}{\partial \xi_1^{m-1}}
\end{bmatrix}(f)
= \mathcal{H}_{pp}(f) \cdot
\begin{bmatrix}
\hat{\mathcal{K}}^{x,m-1}_{p}\\
\hat{\mathcal{K}}^{y,m-1}_{p}\\
\hat{\mathcal{K}}^{z,m-1}_{p}
\end{bmatrix}(f),p \in P_{\xi_1},
\label{dH_localization}
\end{equation}
where $\hat{H}$ and $\hat{\mathcal{K}}$ are Fourier-transformed magnetic field and equivalent source; $\mathcal{H}_{pp}, p\in P_{\xi_1}$ is the transfer function of the cell of $P_{\xi_1}$ itself. Assuming that an incident field is placed at $p$, then its self transfer function is found as:
\begin{equation}
\mathcal{H}_{pp}(f) = \dfrac{\mathfrak{F} [ E^{\text{total}}_p(t)]}{\mathfrak{F} [E^{\text{incident}}_p(t)]}.
\end{equation}
Similarly, the $(m-1)$-th order electric field derivatives at the $P_{\xi_1}$ cells are derived as: 
\begin{equation}
\begin{bmatrix}
\dfrac{\partial^{m-1} \hat{E}^x_{p}}{\partial \xi_1^{m-1}}\\[.3cm]
\dfrac{\partial^{m-1} \hat{E}^y_{p}}{\partial \xi_1^{m-1}}\\[.3cm]
\dfrac{\partial^{m-1} \hat{E}^z_{p}}{\partial \xi_1^{m-1}}
\end{bmatrix}(f)
= \mathcal{H}_{pp}(f) \cdot
\begin{bmatrix}
\hat{\mathcal{J}}^{x,m-1}_{p}\\
\hat{\mathcal{J}}^{y,m-1}_{p}\\
\hat{\mathcal{J}}^{z,m-1}_{p}
\end{bmatrix}(f),p \in P_{\xi_1}.
\label{dE_localization}
\end{equation}
Therefore, (\ref{product_term_in_J}) can be further written in the frequency domain as:
\begin{equation}
\begin{split}
 &\dfrac{\partial^{m-q} \textbf{A}_e}{\partial \xi_1^{m-q}} \dfrac{\partial^q \overset{\star}{\hat{\textbf{H}}}(f)}{\partial \xi_1^q} \\&=
\dfrac{\partial^{m-q} \textbf{A}_e}{\partial \xi_1^{m-q}}
\begin{bmatrix}
\begin{bmatrix}
0 \\ 0 \\ \vdots \\ \mathcal{H}_{pp}(f) \\ \vdots
\end{bmatrix}
&
\begin{bmatrix}
0 \\ 0 \\ \vdots \\ \mathcal{H}_{pp}(f) \\ \vdots
\end{bmatrix}
&
\begin{bmatrix}
0 \\ 0 \\ \vdots \\ \mathcal{H}_{pp}(f) \\ \vdots
\end{bmatrix}
\end{bmatrix}^T 
\cdot
\bm{\hat{\mathcal{K}}}^q(f)
\\ &=
\left( \dfrac{\partial^{m-q} \textbf{A}_e}{\partial \xi_1^{m-q}} \right) \overset{\star}{\bm{\mathcal{H}}}(f) \cdot \hat{\bm{\mathcal{K}}}^q(f)
\end{split}
\label{S_star}
\end{equation}

Combining (\ref{dH_localization}), (\ref{dE_localization}), (\ref{S_star}) and (\ref{JmKm}), the equivalent sources up to any order can be computed in the frequency domain as: 
\begin{equation}
\begin{split}
{\hat{\bm{\mathcal{J}}}^{(m)}}(f) &= 
\sum_{q=0}^{m-1}   \dfrac{m!}{(m-q)! q!}  
\left( \dfrac{\partial^{m-q} \textbf{A}_e}{{\partial \xi_1}^{m-q}} \right) 
\dfrac{\partial^q \overset{\star}{\hat{\textbf{H}}}(f)}{\partial \xi_1^q}
\\&= 
\dfrac{\partial^m \textbf{A}_e}{\partial \xi_1^m}\overset{\star}{\hat{\textbf{H}}(f)} 
\\& \quad + 
\sum_{q=1}^{m-1}   \dfrac{m!}{(m-q)! q!}  
\left( \dfrac{\partial^{m-q} \textbf{A}_e}{{\partial \xi_1}^{m-q}} \right) 
\left( \overset{\star}{\bm{\mathcal{H}}(f)} {\hat{\bm{\mathcal{K}}^q}}(f) \right)
\end{split}
\label{Jstarm}
\end{equation}
Meanwhile, the equivalent source for high-order magnetic field derivatives is found as:
\begin{equation}
\begin{split}
{\hat{\bm{\mathcal{K}}}^{(m)}}(f) &= 
\sum_{q=0}^{m-1}  \dfrac{m!}{(m-q)! q!}  
\left( \dfrac{\partial^{m-q} \textbf{A}_h}{{\partial \xi_1}^{m-q}} \right) 
\dfrac{\partial^q \overset{\star}{\hat{\textbf{E}}}(f)}{\partial \xi_1^q}
\\&= 
\dfrac{\partial^m \textbf{A}_h}{\partial \xi_1^m}\overset{\star}{\hat{\textbf{E}}(f)} 
\\&\quad+ 
\sum_{q=0}^{m-1}   \dfrac{m!}{(m-q)! q!}  
\left( \dfrac{\partial^{m-q} \textbf{A}_h}{{\partial \xi_1}^{m-q}} \right) 
\left( \overset{\star}{\bm{\mathcal{H}}(f)} {\hat{\bm{\mathcal{J}}^q}}(f) \right)
\end{split}
\label{Kstarm}
\end{equation}
Therefore, by recursively exploiting equivalent sources of previous orders, only $\bm{\mathcal{J}}^{(1)}$ and $\bm{\mathcal{K}}^{(1)}$ are needed in order to compute equivalent sources of higher orders $\bm{\mathcal{J}}^{(m)}, \bm{\mathcal{K}}^{(m)}\forall m>1$. Here, $\bm{\mathcal{J}}^{(1)}$ and $\bm{\mathcal{K}}^{(1)}$ are referred as the initial states for high-order equivalent sources and are computed on the fly with a single FDTD run.


\subsection{High-order field derivatives}
\begin{figure}[]
	\centering
    \includegraphics[width=8cm]{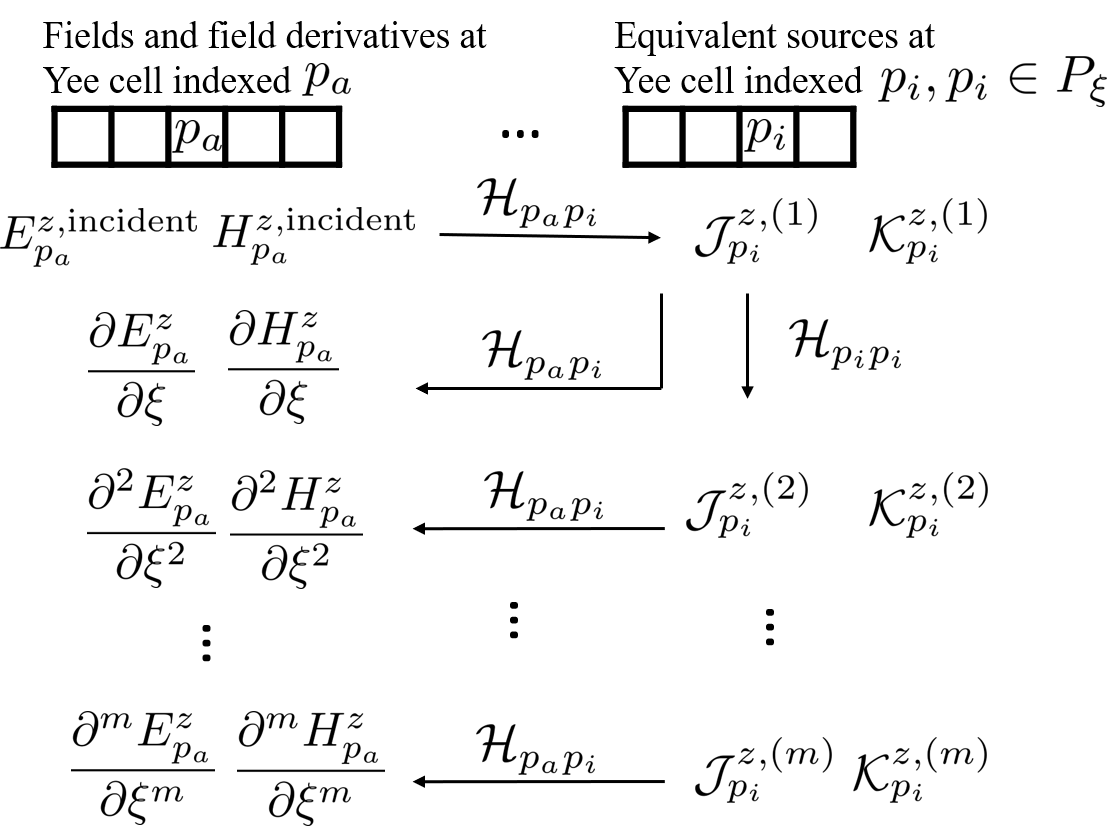}
	\caption{The computation of high-order field derivatives using equivalent sources. The $M$-th order equivalent sources are found by propagating the $(M-1)$-th order counterparts through the transfer function of the cell $p_i$ itself. Afterwards, the field derivatives are computed by propagating the equivalent sources at the cell $p_i$ to the observation point $p_a$.} 
	\label{zigzag}
\end{figure}

Using the equivalent sources for $m$-th order field derivatives with respect to $\xi_1$, i.e., $\bm{\mathcal{J}}^{(m)}$ and $\bm{\mathcal{K}}^{(m)}$, the $m$-th order field derivatives at the observation point indexed $p_a$ are computed using the relation of (\ref{eq_relation1}):
\begin{equation}
\dfrac{\partial^m \hat{E^{z}_{p_a}}}{\partial \xi_1^m}(f) = 
\displaystyle{\sum_{p_i \in P_{\xi_1}}} \mathcal{H}_{p_i p_a}(f)\cdot
\hat{\mathcal{J}}^{z,(m)}_{p_i}(f).
\label{highorder_eq}
\end{equation}
In (\ref{highorder_eq}), the $m$-th order field derivative is found by propagating the equivalent sources from the cells of $P_{\xi_1}$ to the observation point through transfer functions between them.  

Fig. \ref{zigzag} illustrates the computation of high-order field derivatives by using this proposed method. In addition to transfer functions between observation point and $P_{\xi_1}$ cells, the transfer functions of the $P_{\xi_1}$ cells themselves are used to compute high-order equivalent sources. 

Based on this method, only two FDTD simulations are necessary in order to compute these transfer functions and the initial state of equivalent sources. The first FDTD simulation is run with Gaussian excitation placed at the cells of $P_{\xi_1}$ to obtain $\mathcal{H}_{p_ip_i}, p_i \in P_{\xi_1}$. Afterwards, the second FDTD simulation is run with a Gaussian excitation placed at $p_a$. After the grid is excited, the transfer function $\mathcal{H}_{p_ip_a}$ as well as $\bm{\mathcal{J}}^{(1)}$, $\bm{\mathcal{K}}^{(1)}$ are derived following the procedure stated in Algorithm 1. The equivalent sources for first order field derivatives, $\bm{\mathcal{J}}^{(1)}$ and $\bm{\mathcal{K}}^{(1)}$, are the fundamental terms to compute higher-order equivalent sources. 

Algorithm 2 summarizes the procedure, including the two FDTD runs and the post-processing part. The first FDTD run is used to find the transfer functions at the subset of affected cells and the second is used to compute the initial state of equivalent sources as well as the field solution to the original problem. In the post-processing part, high-order equivalent sources are propagated to the observation port through the transfer function.

\begin{algorithm}
 \caption{Compute High-order Field Derivatives with Equivalent Sources in FDTD}
 \begin{algorithmic}[1]
  \Statex $P_\xi$: the subset of the cells affected by parameters $\xi$
  \Statex $p_a$: observation port 
  \Statex $p$: matrix index mapping triad $(i,j,k)$, $p = 1,2,\dots,P$
\\ \textbf{procedure} \textsc{Self-Transfer Functions of $P_\xi$ Cells}
   \Indent
	\FOR {$t = 1$ to $Tsteps$}
  	\IF {$t<\text{excitation time}$}
  		\STATE $E^z_{p\in P_\xi} \gets g(t)$ Gaussian excitation
  	\ENDIF
  	\STATE Update $E^{x,y,z}(1:P)$
  	\STATE Update $H^{x,y,z}(1:P)$
  \ENDFOR 
  \STATE Compute $\mathcal{H}_{pp}(f)\gets \mathfrak{F}[E^{\text{total}}_p(t)]/\mathfrak{F}[E^{\text{inc}}_p(t)], p \in P_{\xi}$
    \EndIndent
\\ \textbf{procedure} \textsc{FDTD Run}
\Statex $A^{x,y,z}_{e,h}(p)'$: first-order derivatives of update coefficients w.r.t $\xi$
\Indent
  \FOR {$t = 1$ to $Tsteps$}
  	\IF {$t<\text{excitation time}$}
  		\STATE $E^z_{p_a} \gets g(t)$ Gaussian excitation
  	\ENDIF
  	\STATE Update $E^{x,y,z}(1:P)$
  	\STATE Update $H^{x,y,z}(1:P)$
  	\STATE Record fields at $p\in P_{\xi}$ 
  	\STATE Compute $\mathcal{J}^{x,y,z}$ and $\mathcal{K}^{x,y,z}$ using (\ref{j_first_order}):
  	\Indent
	  	\STATE $\mathcal{J}^{x,y,z}(1:P) \gets {A^{x,y,z}_e(1:P)}', H^{x,y,z}(1:P) $ 
	  	\STATE $\mathcal{K}^{x,y,z}(1:P) \gets {A^{x,y,z}_h(1:P)}', E^{x,y,z}(1:P) $
	\EndIndent 
  \ENDFOR   
\EndIndent
\\ \textbf{procedure} \textsc{Compute $M$-th order Equivalent Sources}
\Indent
	\STATE $\mathcal{J}^1(f)\gets \mathfrak{F}[\mathcal{J}(t)]$ and $\mathcal{K}^1(f)\gets \mathfrak{F}[\mathcal{K}(t)]$
	\STATE $H_p(f)\gets \mathfrak{F}[H_p(t)]$, $E_p(f)\gets \mathfrak{F}[E_p(t)]$, $p\in P_\xi$
	\STATE Use (\ref{JmKm}) to compute $\mathcal{J}^m_p,\mathcal{K}^m_p, p \in P_{\xi}$ in freq. domain:
	\FOR {$m = 2$ to $M$}
	\STATE $\mathcal{J}^{m}_p (f)\gets \frac{\partial^m A_e}{\partial \xi^m} H_p(f) + \sum\limits_{q=1}^{m-1} \binom{m-1}{q} \mathcal{H}_{pp}(f) \mathcal{K}^{q}(f)$
	\STATE $\mathcal{K}^{m}_p (f)\gets \frac{\partial^m A_h}{\partial \xi^m} E_p(f) + \sum\limits_{q=1}^{m-1} \binom{m-1}{q} \mathcal{H}_{pp}(f) \mathcal{J}^{q}(f)$
	\ENDFOR
\EndIndent

\\ \textbf{procedure} \textsc{High-order Equivalent Source Propagation}
\Indent
\STATE Find the transfer function using (\ref{transfer_fnt})
     \Indent
     \STATE $\mathcal{H}_{p_ap}(f) \gets \mathfrak{F}[E^z_{p}(t)]/ \mathfrak{F}[E^z_{p_a}(t)], p \in P_{\xi}$
     \EndIndent
\STATE Propagate $m$-th order equivalent source using (\ref{highorder_eq})
\Indent
\STATE $\frac{\partial^m E^z_{p_a}}{\partial \xi^m}(f) \gets \mathcal{H}_{p_ap}(f) \cdot \mathcal{J}^{z,m}_{p}(f), p \in P_\xi$
\EndIndent
\EndIndent
 \end{algorithmic} 
 \end{algorithm}

\subsection{Partial mixed derivatives and Hessian matrix} 
In addition to the Jacobian matrix and high-order field derivatives, the Hessian matrix is another important part of sensitivity analysis and optimization in electromagnetic design. Consider the Hessian matrix of $E^{z}_{p_a}$, which is the $z$-component of the electric field at port $p_a$, with respect to $N$ parameters $\bm{\xi}=[\xi_1, \xi_2, \dots, \xi_N]^T$:
\begin{equation}
\mathbb{H}(E^{z}_{p_a})\equiv
\begin{bmatrix}
    \dfrac{{\partial}^2 E^z_{p_a} }{\partial {\xi_1}^2} & \dfrac{{\partial}^2 E^z_{p_a} }{\partial \xi_1 \partial \xi_2} & \dots & \dfrac{{\partial}^2 E^z_{p_a} }{\partial \xi_1 \partial \xi_N}   
 \\[.4cm]
    \dfrac{{\partial}^2 E^z_{p_a} }{\partial \xi_2 \partial \xi_1} & \dfrac{{\partial}^2 E^z_{p_a} }{\partial {\xi_2}^2 } & \dots & \dfrac{{\partial}^2 E^z_{p_a} }{\partial \xi_2 \partial \xi_N} 
\\
\vdots & \vdots & \ddots & \vdots
\\
\dfrac{{\partial}^2 E^z_{p_a} }{\partial \xi_N \partial \xi_1} & \dfrac{{\partial}^2 E^z_{p_a} }{\partial \xi_N \partial \xi_2 } & \dots & \dfrac{{\partial}^2 E^z_{p_a} }{\partial {\xi_N}^2} 
\end{bmatrix}
.
\end{equation}
The diagonal of this Hessian matrix consists of second-order field derivatives with respect to $\xi_1, \xi_2, \dots, \xi_N$, respectively. These second-order field derivatives are computed using the method proposed in the previous section. In addition, the upper and lower triangular parts of this matrix are the partial mixed derivatives with respect to any two variables out of $\bm{\xi}$.

Direct differentiation is performed to (\ref{fdtd_eq}) with respect to $\xi_u$ and $\xi_v$, $u \neq v$, yielding:
\begin{align}
&\begin{cases}
\dfrac{\partial^2 \textbf{E}^{n+1}}{\partial \xi_u \partial \xi_v} = 
\dfrac{\partial^2 \textbf{E}^{n}}{\partial \xi_u \partial \xi_v} + 
\textbf{A}_e \dfrac{\partial^2 \textbf{H}^{n+\frac{1}{2}}}{\partial \xi_u \partial \xi_v} + 
\bm{\mathcal{J}}^{uv}
\\[.2cm]
\dfrac{\partial^2 \textbf{H}^{n+\frac{1}{2}}}{\partial \xi_u \partial \xi_v} = 
\dfrac{\partial^2 \textbf{H}^{n-\frac{1}{2}}}{\partial \xi_u \partial \xi_v} + 
\textbf{A}_h \dfrac{\partial^2 \textbf{E}^{n}}{\partial \xi_u \partial \xi_v} + 
\bm{\mathcal{K}}^{uv}
\end{cases}
\label{diff_fdtd_cross}
\end{align}
\begin{multline}
\bm{\mathcal{J}}^{uv}
 =  \dfrac{\partial \textbf{A}_e}{\partial \xi_u} \dfrac{\partial \textbf{H}^{n+\frac{1}{2}}}{\partial \xi_v} 
+
\dfrac{\partial \textbf{A}_e}{\partial \xi_v} \dfrac{\partial \textbf{H}^{n+\frac{1}{2}}}{\partial \xi_u} + 
\cancelto{0}{
\dfrac{\partial^2 \textbf{A}_e}{\partial \xi_u \partial \xi_v}}
\textbf{H}^{n+\frac{1}{2}},
\\
\bm{\mathcal{K}}^{uv}
 =  \dfrac{\partial \textbf{A}_h}{\partial \xi_u} \dfrac{\partial \textbf{E}^{n}}{\partial \xi_v} 
+
\dfrac{\partial \textbf{A}_h}{\partial \xi_v} \dfrac{\partial \textbf{E}^{n}}{\partial \xi_u} + 
\cancelto{0}{
\dfrac{\partial^2 \textbf{A}_h}{\partial \xi_u \partial \xi_v}}
\textbf{H}^{n}.
 \label{J_cross}
\end{multline}
Eq. (\ref{diff_fdtd_cross}) includes the FDTD update equations with respect to partial mixed field derivatives. Evidently, the partial mixed field derivatives are excited by $\bm{\mathcal{J}}^{uv}$ and $\bm{\mathcal{K}}^{uv}$. 
Assuming that the perturbations of $\xi_u$ and $\xi_v$ are mapped to the $d^y$ of the cells in the subset $P_{\xi_u}$ and $P_{\xi_v}$ respectively, and $P_{\xi_u} \cap P_{\xi_v} = \emptyset$. Therefore, the $\frac{\partial^2 \textbf{A}_{e}}{\partial \xi_u \partial \xi_v}$ and $\frac{\partial^2 \textbf{A}_{h}}{\partial \xi_u \partial \xi_v}$ terms in (\ref{J_cross}) are assumed to be zero.

As discussed in the previous section, in the derivatives of update matrices, only the terms corresponding to the cells of $P_{\xi_u}$ and $P_{\xi_v}$ are non-zero. Specifically, in (\ref{J_cross}):
\begin{equation}
\begin{split}
\dfrac{\partial \textbf{A}_e}{\partial \xi_u} = \Delta t \textbf{M}^{-1}_{\epsilon}
\left(
\textbf{C}_{e} \circ 
\begin{bmatrix}
\bm{0} & \bm{0} & \partial \textbf{D}^y/\partial {\xi_u} \\
\bm{0} & \bm{0} &\bm{0} \\
\partial \textbf{D}^y/\partial {\xi_u} & \bm{0} & \bm{0} 
\end{bmatrix}
\right)
\\
\dfrac{\partial \textbf{A}_e}{\partial \xi_v} = \Delta t \textbf{M}^{-1}_{\epsilon}
\left(
\textbf{C}_{e} \circ 
\begin{bmatrix}
\bm{0} & \bm{0} & \partial \textbf{D}^y/\partial {\xi_v} \\
\bm{0} & \bm{0} &\bm{0} \\
\partial \textbf{D}^y/\partial {\xi_v} & \bm{0} & \bm{0} 
\end{bmatrix}
\right)
\end{split},
\end{equation}
where
\begin{equation}
\begin{split}
\dfrac{\partial \textbf{D}^y}{\partial {\xi_u}}=  \dfrac{ \partial \left[ D^y \right]_{ij}}{\partial {\xi_u}}=
\begin{cases}
\dfrac{\partial (1/\tilde{\Delta y_i})}{\partial \xi_u}, i = j \in P_{\xi_u}
\\
0, \text{otherwirse}.
\end{cases}
\\
\dfrac{\partial \textbf{D}^y}{\partial {\xi_v}}= \dfrac{\partial \left[ D^y \right]_{ij}}{\partial {\xi_v}}=
\begin{cases}
\dfrac{\partial (1/\tilde{\Delta y_i})}{\partial \xi_v}, i = j \in P_{\xi_v}
\\
0, \text{otherwise}.
\end{cases}
\end{split}
\end{equation}
Therefore, the product of the field derivative with respect to $\xi_u$ and the derivative of the update matrix with respect to $\xi_v$ is localized at the cells of $P_{\xi_u}$:
\begin{equation}
\begin{split}
\dfrac{\partial \textbf{A}_e}{\partial \xi_u} \dfrac{\partial \textbf{H}}{\partial \xi_v} 
&= 
\dfrac{\partial \textbf{A}_e}{\partial \xi_u} \dfrac{\partial \overset{\star}{\textbf{H}}}{\partial \xi_v}
\\ &\equiv
\dfrac{\partial \textbf{A}_e}{\partial \xi_u}
\begin{bmatrix}
\begin{bmatrix}
0 \\ 0 \\ \vdots \\ \frac{\partial H^x_{p}}{\partial \xi_v} \\ \vdots
\end{bmatrix}
&
\begin{bmatrix}
0 \\ 0 \\ \vdots \\ \frac{\partial H^y_{p}}{\partial \xi_v} \\ \vdots
\end{bmatrix}
&
\begin{bmatrix}
0 \\ 0 \\ \vdots \\ \frac{\partial H^z_{p}}{\partial \xi_v} \\ \vdots
\end{bmatrix}
\end{bmatrix}^T,
p \in P_{\xi_u}
\end{split}.
\label{anchor}
\end{equation}
Notably, the first term in (\ref{anchor}), which is the derivative of the update matrix with respect to $\xi_u$ or $\xi_v$ anchors the field derivatives needed at the cells of $P_{\xi_u}$ or $P_{\xi_v}$, which are then used to compute the equivalent sources. Substituting (\ref{anchor}) into (\ref{J_cross}), $\bm{\mathcal{J}}^{uv}$ is found in frequency domain as:
\begin{multline}
{\hat{\bm{\mathcal{J}}}^{uv}}(f) = 
\sum_{p_v \in P_{\xi_v}} \sum_{p_u \in P_{\xi_u}}
\\
\left(
\dfrac{\partial \textbf{A}_e}{\partial \xi_u}
[{{\mathcal{H}}_{p_vp_u}} (f) \hat{\bm{\mathcal{K}}}^{v}(f) ]
+ \dfrac{\partial \textbf{A}_e}{\partial \xi_v}
[{{\mathcal{H}}_{p_up_v}}(f) \hat{\bm{\mathcal{K}}}^{u} (f) ]
\right),
\label{Juv}
\end{multline}
where $\bm{\mathcal{K}}^{u}$ and $\bm{\mathcal{K}}^{v}$ are the equivalent sources for the magnetic field derivative with respect to $\xi_u$ and $\xi_v$ respectively; $\mathcal{H}_{p_up_v}(f)$ is the transfer functions between two subsets of affected cells. Similarly,
\begin{multline}
{\hat{\bm{\mathcal{K}}}^{uv}}(f) = 
\sum_{p_v \in P_{\xi_v}} \sum_{p_u \in P_{\xi_u}}
\\
\left(
\dfrac{\partial \textbf{A}_h}{\partial \xi_u}
[{{\mathcal{H}}_{p_vp_u}} (f) \hat{\bm{\mathcal{J}}}^{v}(f) ]
+ \dfrac{\partial \textbf{A}_h}{\partial \xi_v}
[{{\mathcal{H}}_{p_up_v}}(f) \hat{\bm{\mathcal{J}}}^{u} (f) ]
\right).
\label{Kuv}
\end{multline}
Eq. (\ref{Juv}) and (\ref{Kuv}) indicate that the equivalent sources for the partial mixed field derivatives with respect to $\xi_u$ and $\xi_v$, are localized at the cells of subsets $P_{\xi_u}$ and $P_{\xi_v}$. For example,
\begin{equation}
\bm{\mathcal{J}}^{uv}=
\begin{cases}
\left\lbrace \mathcal{J}^{x,uv}_p, \mathcal{J}^{y,uv}_p, \mathcal{J}^{z,uv}_p\right\rbrace, p \in P_{\xi_u} or P_{\xi_v},
\\
\left\lbrace 0,0,0 \right\rbrace \text{otherwise}.
\end{cases}
\end{equation}

Using the relation stated in (\ref{eq_relation1}), the partial mixed field derivative with respect to $\xi_u$ and $\xi_v$, at the port $p_a$ is found by propagating the equivalent source from the cells of $P_{\xi_u}$ and $P_{\xi_v}$ to $p_a$:
\begin{multline}
\dfrac{\partial^2 \hat{E^z_{p_a}}}{\partial \xi_u \partial \xi_v}(f) = 
\sum_{p_u \in P_{\xi_u}} 
\mathcal{H}_{p_up_a}(f)\cdot \hat{\mathcal{J}^{uv}_{p_u}}(f) 
\\+ 
\sum_{p_v \in P_{\xi_v}}
\mathcal{H}_{p_vp_a}(f) \cdot \hat{\mathcal{J}^{uv}_{p_v}}(f).
\label{JuvPropagte}
\end{multline}

Algorithm 3 summarizes the computation of the partial-mixed field derivatives  using equivalent sources. For $N$ design parameters under consideration, a total of $N+1$ FDTD runs are needed. The $N$ runs are used to compute the transfer functions between the affected subsets corresponding to each parameter. The last FDTD run is used to compute the equivalent sources. Finally, these equivalent sources recorded at different subsets are propagated to the observation point through transfer functions.

\begin{algorithm}
 \caption{Compute Partial Mixed Field Derivatives with Equivalent Sources in FDTD }
 \begin{algorithmic}[1]
 \Statex $P_1,P_2,...,P_N$: subsets of the cells affected by parameters $   \xi_1, \xi_2, ...\xi_N$
  \Statex $p_a$: observation point
  \Statex $p$: matrix index mapping triad $(i,j,k)$, $p = 1,2,\dots,P$
\\ \textbf{procedure} \textsc{System Analysis}
\Indent
	\FOR {$u = 1$ to $N$}
	\STATE Set excitation $E^z_{p\in P_{u}} \gets g(t)$ Gaussian excitation
	\STATE Run FDTD 
  	\STATE Record transmitted fields: $E^{\text{trans}}_{p\in{P_u}}(t)$
     \FOR {$ v = 1$ to $N$, $v \neq u$}
    \STATE $\mathcal{H}_{p_up_v}(f) \gets \mathfrak{F}[E^{\text{trans}}_{p_v \in{P_v}}]/\mathfrak{F}[E^{\text{inc}}_{p_u \in{P_u}}]$
    \ENDFOR
    \ENDFOR
\EndIndent
    
\\ \textbf{procedure} \textsc{Perturbation Mapping}
\Indent
  \FOR {$p = 1$ to $P$}
    \IF {$p \in P_{n}, n = 1:N$}
  \STATE ${A^{x,y,z}_e(p)}' \gets \frac{\partial A^{x,y,z}_e(p)}{\partial \xi_n} $
  \STATE ${A^{x,y,z}_h(p)}' \gets \frac{\partial A^{x,y,z}_h(p)}{\partial \xi_n} $
     \ENDIF
  \ENDFOR
\EndIndent

\\ \textbf{procedure} \textsc{FDTD Run}
\Indent
  \FOR {$t = 1$ to $Tsteps$}
  	\IF {$t<\text{excitation time}$}
  		\STATE $E^z_{p_a} \gets g(t)$ Gaussian excitation
  	\ENDIF
  	\STATE Update $E^{x,y,z}(1:P)$
  	\STATE Update $H^{x,y,z}(1:P)$
  	\STATE Compute $\mathcal{J}^{x,y,z}$ and $\mathcal{K}^{x,y,z}$ using (\ref{j_first_order}):
  	\Indent
	  	\STATE $\mathcal{J}^{x,y,z}(1:P) \gets {A^{x,y,z}_e(1:P)}', H^{x,y,z}(1:P) $ 
	  	\STATE $\mathcal{K}^{x,y,z}(1:P) \gets {A^{x,y,z}_h(1:P)}', E^{x,y,z}(1:P) $
	\EndIndent
	\FOR{$n=1$ to $N$}
	\IF {$ p \in P_n$}
	  \STATE $\mathcal{J}^n(t),\mathcal{K}^n(t) = \mathcal{J}_p(t),\mathcal{K}_p(t)$
	\ENDIF
	\ENDFOR 
  \ENDFOR   
\EndIndent

\\ \textbf{procedure} \textsc{Post-Processing}
\Statex $\mathcal{J}^{uv},\mathcal{K}^{uv}$: Equivalent sources for partial mixed field derivatives w.r.t $\xi_u, \xi_v$
\Indent
    \FOR {$u=1$ to $N$}
    \FOR {$v=1$ to $N$}
     \STATE Compute $\mathcal{J}^{uv},\mathcal{K}^{uv}$ using (\ref{Juv}) and (\ref{Kuv}):
     \STATE $\mathcal{J}^{uv}(f) \gets \mathfrak{F}[\mathcal{K}^{u}(t)], \mathfrak{F}[\mathcal{K}^{v}(t)], \mathcal{H}_{p_ip_j, p_i \in P_u, p_j \in P_v}$
     \STATE $\mathcal{K}^{uv}(f) \gets \mathfrak{F}[\mathcal{J}^{u}(t)], \mathfrak{F}[\mathcal{J}^{v}(t)], \mathcal{H}_{p_ip_j, p_i \in P_u, p_j \in P_v}$
     \STATE Propagate the equivalent source using (\ref{JuvPropagte}):
     \STATE $\mathcal{H}_{p_ap_{i}}(f) \gets \mathfrak{F}[E^z_{p_{i}\in P_u}(t)]/ \mathfrak{F}[E^z_{p_a}(t)]$
     \STATE $\frac{\partial^2 E^z_{p_a}}{\partial \xi_u \partial \xi_v}(f) \gets \mathcal{H}_{p_ap_i}(f), \mathcal{J}^{uv}(f)$
    \ENDFOR
    \ENDFOR
\EndIndent
 \end{algorithmic} 
 \end{algorithm}

\section{Numerical Results}
\label{numericalpart}
Two numerical examples, including a dielectric multilayer and a three-stub microstrip line are presented in this section to demonstrate using the proposed methods to compute Jacobian matrix, high-order and partial mixed field derivatives. The results computed by this paper are marked proposed, while results computed by the centered finite-difference method are marked CFD-FDTD.

\subsection{Dielectric multilayer}
\begin{figure}[h]
	\centering
    \includegraphics[width=9cm]{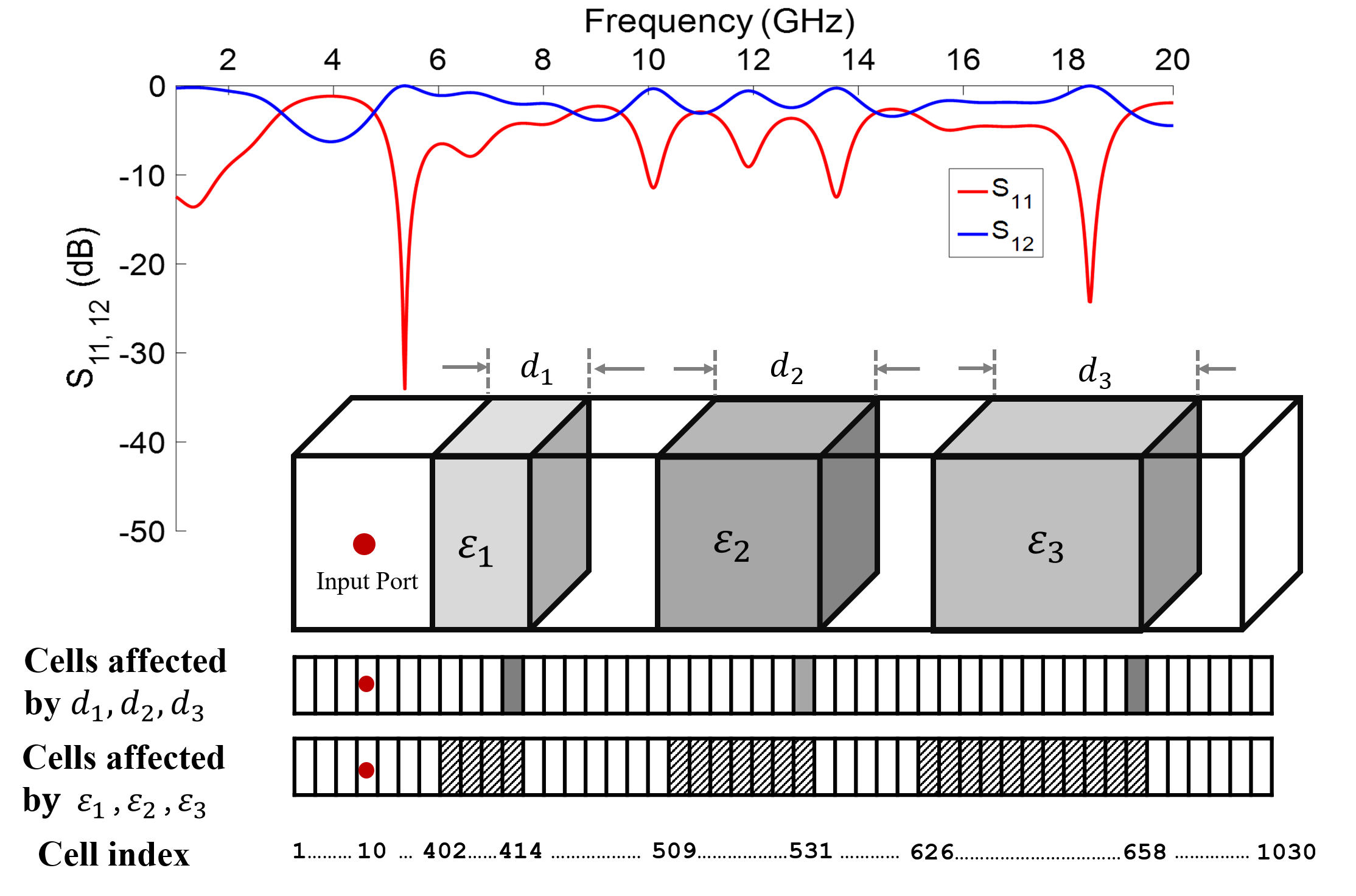}
	\caption{The simulated dielectric multilayer structure. The sensitivity of the fields with respect to six design parameters, including $d_1, d_2, d_3$ and $\epsilon_{r1}, \epsilon_{r2}, \epsilon_{r3}$ are derived.} 
	\label{example1}
\end{figure}

The first example is a 1-D dielectric multilayer geometry consisting of three polycarbonate slabs with $d_1 = 0.0051, d_2 = 0.0093, d_3 = 0.0136$ m, shown in Fig. \ref{example1}. The relative permittivity of each slab is $\epsilon_{r1} = 2.2, \epsilon_{r2} = 3.0, \epsilon_{r3} = 4.0$. Two $0.025$ m air-gaps are located between the slabs. The structure is discretized with $1030$ cells. The cell size $\Delta z= 0.424$mm. $4096$ time steps ($\Delta t = 0.441 $ps) are run and a Gaussian pulse $g(t) = \exp (-(t-t_0)^2/T_s^2)$, with $t_0=6 Ts$ and $T_s=20$ ps is used to excite the grid. The reflection and transmission coefficient $S_{11,12}$ are shown in Fig. \ref{example1} as well.

\begin{figure*}[b]
	\centering
	\vspace{-0.6cm}
    \subfigure[]{
    \includegraphics[height=4.4cm,width=5cm]{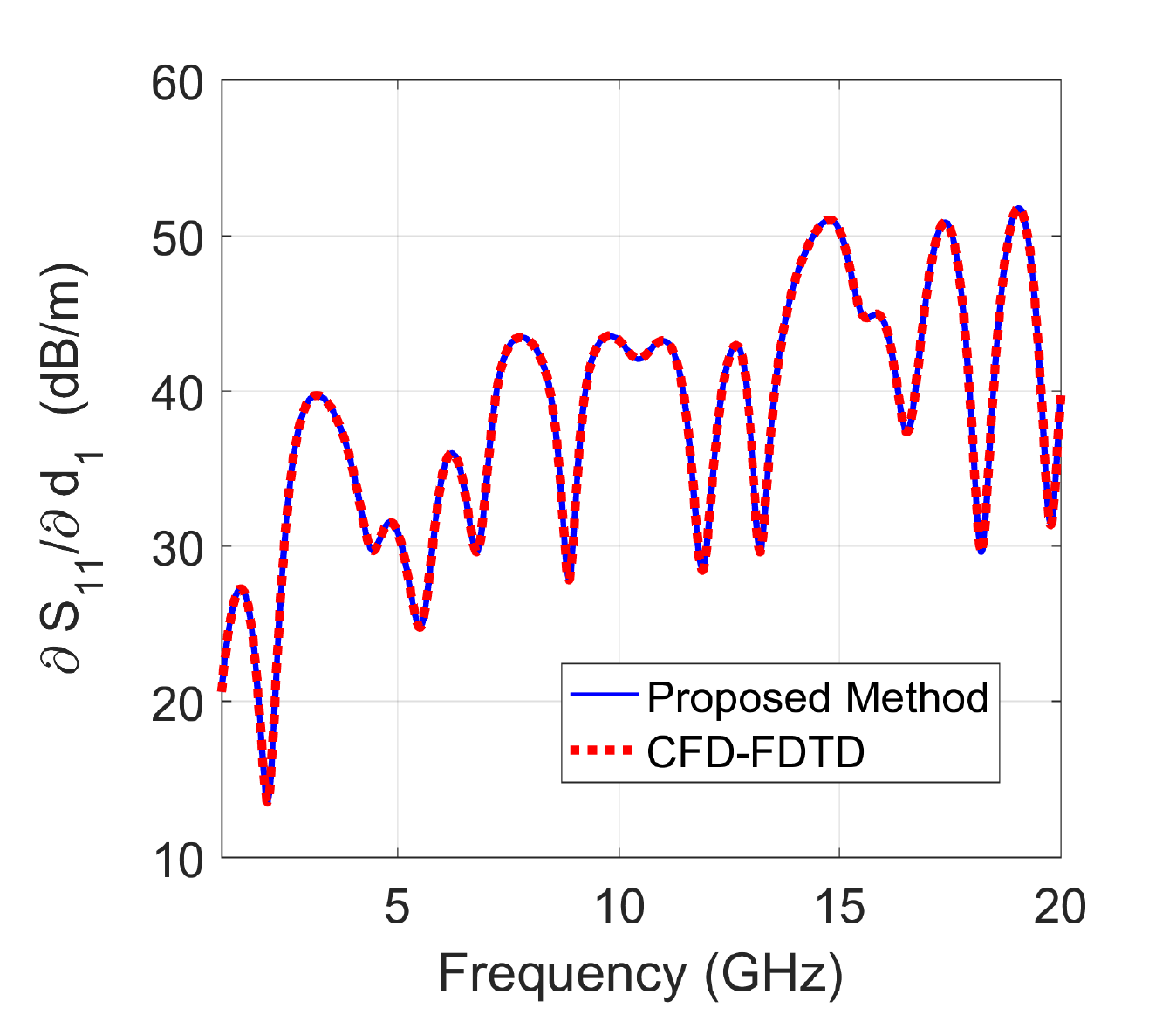}
    \hspace{1.05cm}
    \includegraphics[height=4.4cm,width=5cm]{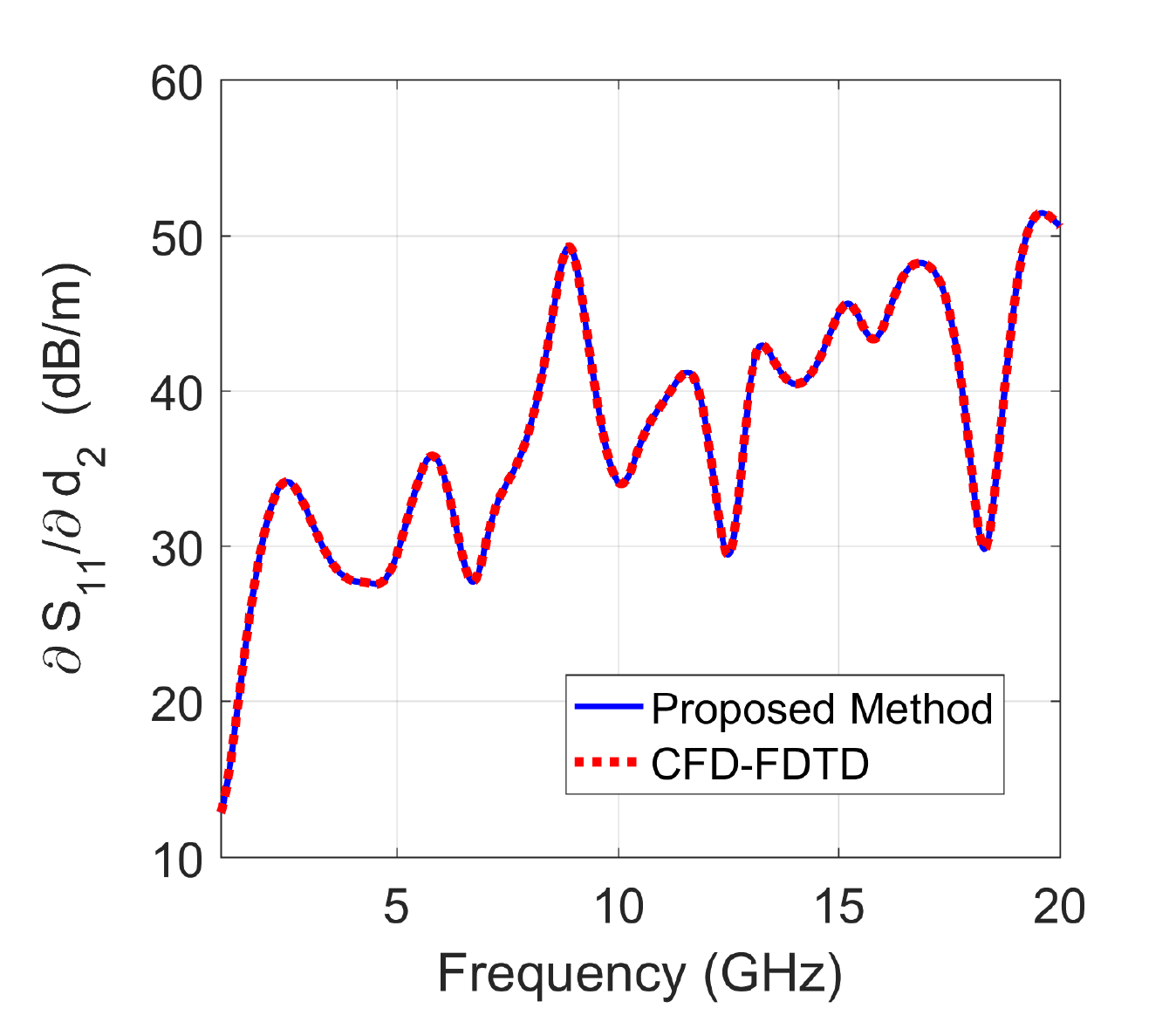}
    \hspace{1.05cm}
    \includegraphics[height=4.4cm,width=5cm]{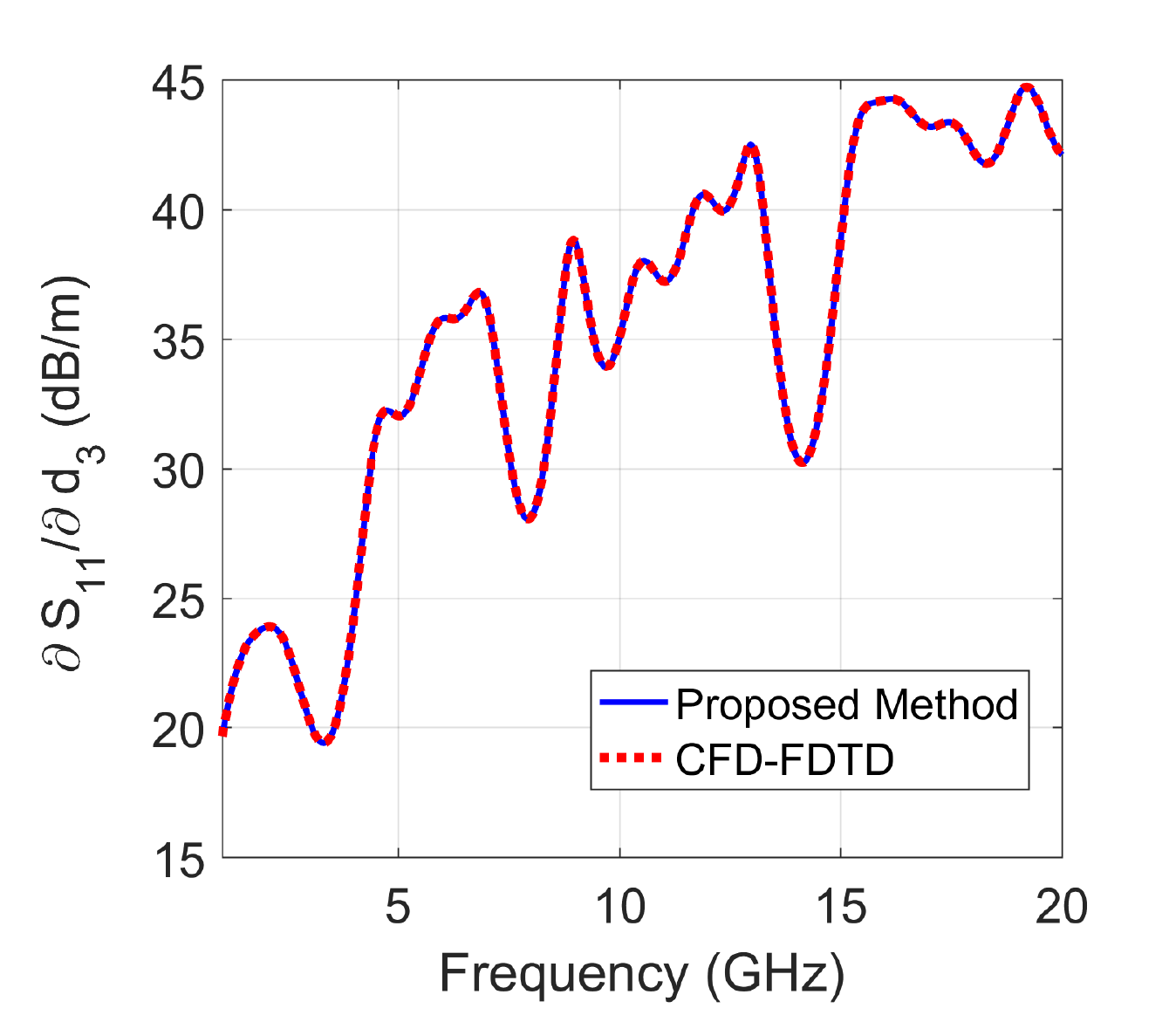}}
    \subfigure[]{ 
    \includegraphics[height=4.4cm,width=5cm]{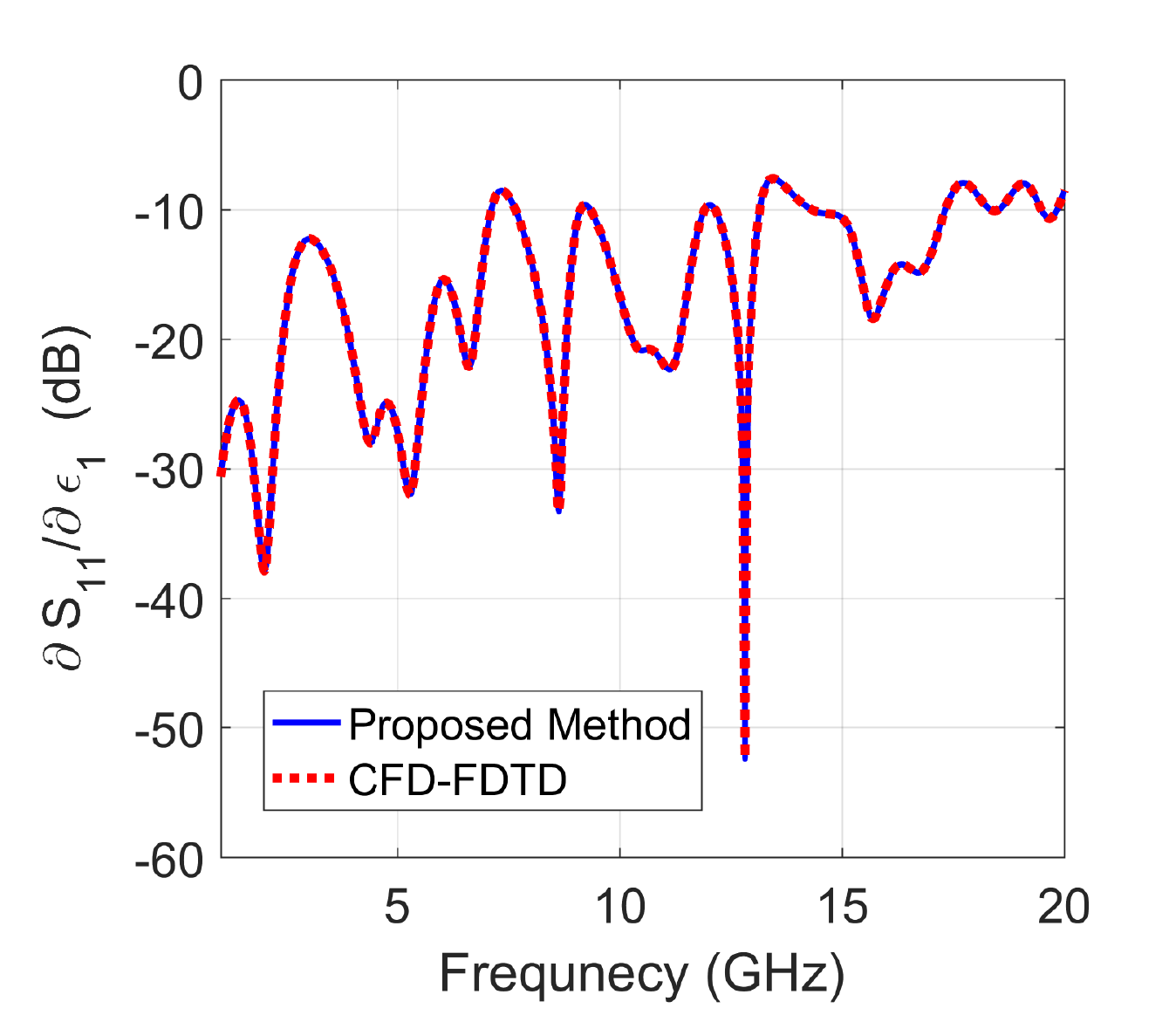}
    \hspace{1.05cm}
    \includegraphics[height=4.4cm,width=5cm]{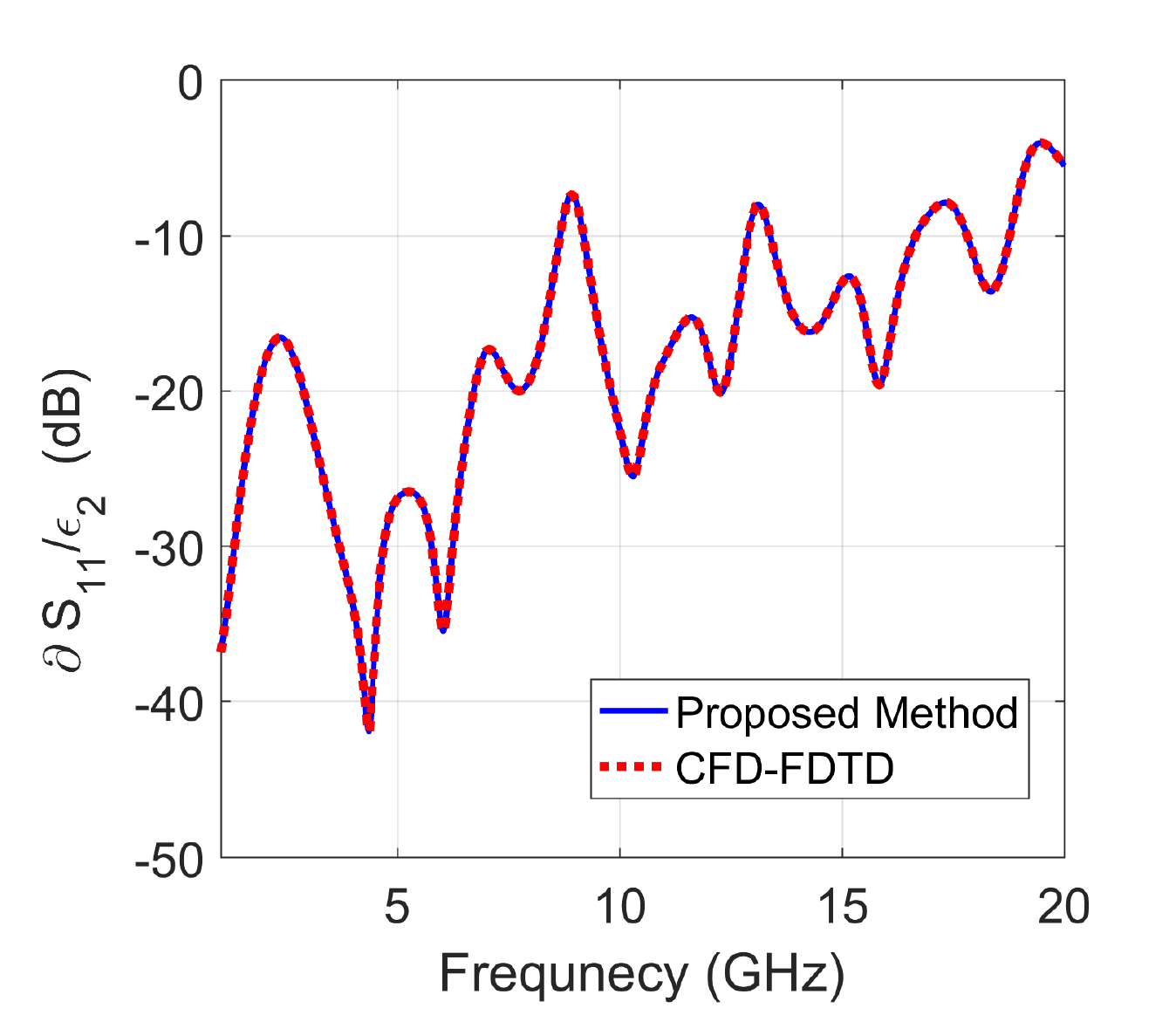}
    \hspace{1.05cm}
    \includegraphics[height=4.4cm,width=5cm]{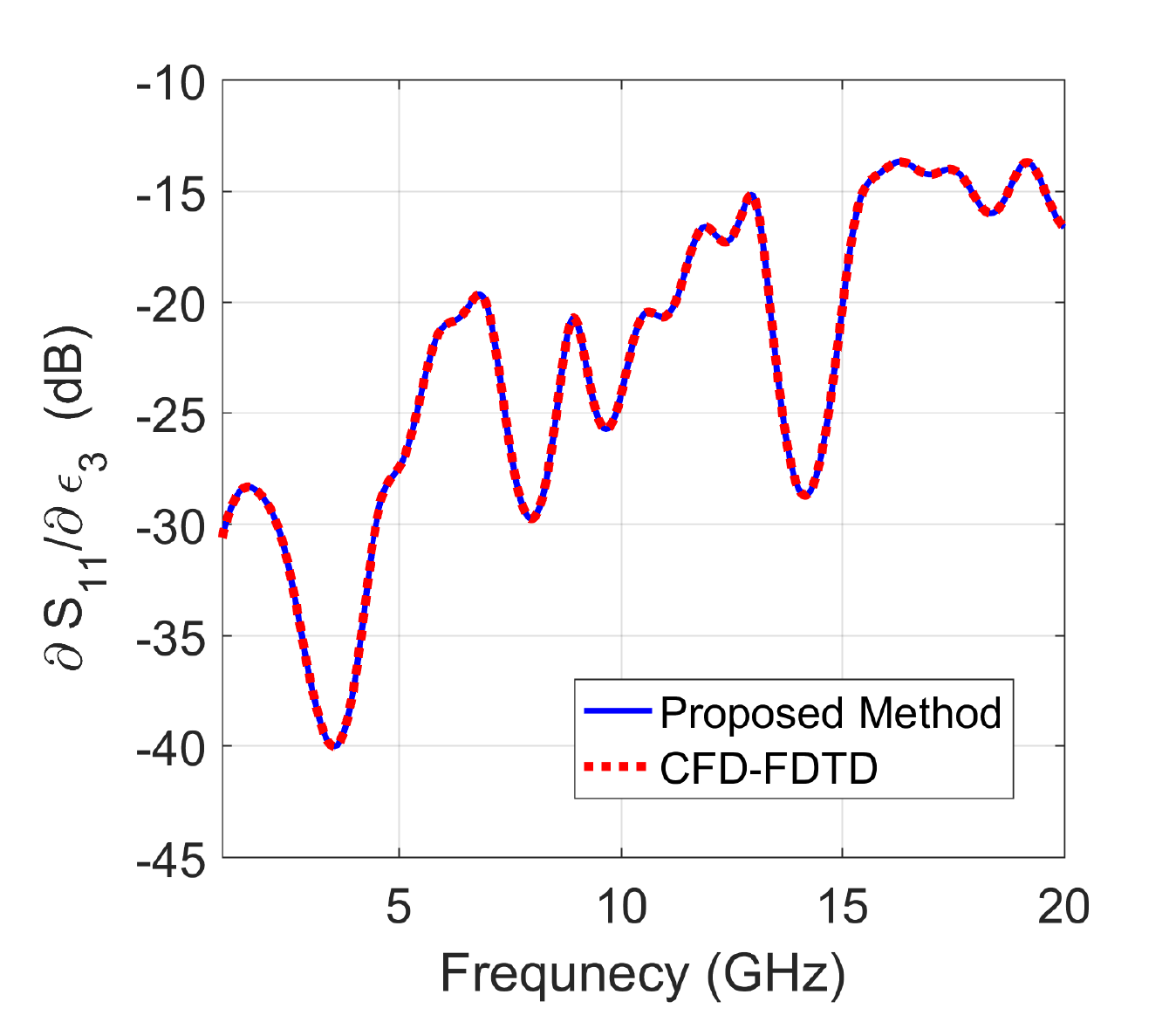}
    \vspace{-0.3cm}}
    \caption{Comparison of the (a) geometric and (b) material sensitivities of $S_{11}$ computed by the proposed method and CFD-FDTD method.}
	\label{slab_jacobian_d}
\end{figure*}

\subsubsection{Jacobian matrix of the reflection coefficient} 
The Algorithm 1 proposed in Section \ref{Multi-parametric} is used to compute the first-order field derivatives with respect to design parameters $d_{1,2,3}$ and $\epsilon_{r1,r2,r2}$. In this example, the derivatives of the reflected fields observed at the input port of this structure are computed. Hence, the derivatives of the reflection coefficient $S_{11}$ with respect to these parameters can be found by applying the chain rule.

As illustrated in the bottom part of Fig. \ref{example1}, the input port is at the cell indexed $10$. The three dielectric slabs are distributed over the cells indexed ${(402:414), (509:531), (626:658)}$ respectively.The perturbations of $d_1, d_2, d_3$ affect the size of the cells indexed ${414, 531 ,658}$, corresponding to the edge of each slab. These cells are coloured grey in Fig. \ref{example1}. Following the Algorithm 1, the Gaussian excitation is placed at cell ${10}$ to excite the grid. The incident fields and the fields probed at cells ${414, 531, 658}$ are used to compute the transfer functions between the affected cells and the input port. While updating fields in the main FDTD simulation, the equivalent sources $\mathcal{J}$ and $\mathcal{K}$ are computed in parallel using (\ref{J_DX}). The equivalent sources for the field derivative with respect to $d_{1,2,3}$ are found by probing $\mathcal{J}$ at cells ${414, 531, 658}$. That is, at each time step, $\mathcal{J}_{d_1}(t) = \mathcal{J}(414)$ $\mathcal{J}_{d_2}(t) = \mathcal{J}(531)$ and $\mathcal{J}_{d_3}(t) = \mathcal{J}(658)$. 

In post-processing, using (\ref{eq_relation1}), the field derivatives at the input port with respect to $d_1$ are computed by the product of the transfer function between cells ${10}$ and ${414}$, and $\mathcal{J}_{d_1}(f)$ in the frequency domain. This post-processing step is repeated to find the field derivatives with respect to other design parameters as well. The derivatives of $S_{11}$ with respect to $d_1, d_2, d_3$ are reported in Fig. \ref{slab_jacobian_d} (a). The results computed by using CFD-FDTD are also presented to verify the correctness of this proposed method.

Similarly, the field derivatives with respect to material parameters $\epsilon_{r1,r2,r2}$ can be computed in parallel. In Fig. \ref{example1}, the cells affected by the perturbations of  $\epsilon_{r1,r2,r2}$ are filled with slash lines, forming subsets $P_{\epsilon_{r1}}, P_{\epsilon_{r2}}, P_{\epsilon_{r3}}$.  Evidently, the equivalent sources for field derivatives with respect to material parameters, are distributed over each slab. Again, (\ref{eq_relation1}) is used to compute the field derivatives as the equivalent sources distribute over a volume. Fig. \ref{slab_jacobian_d} (b) presents the derivative of $S_{11}$ with respect to $\epsilon_{r1,r2,r3}$. In summary, Fig. \ref{slab_jacobian_d} is the Jacobian matrix of the reflection coefficient of this structure over a wide frequency range. Notably, all the information needed to compute this matrix is available with one FDTD simulation, while 12 FDTD simulations are needed to compute this matrix via CFD-FDTD, excluding the runs to ensure the convergence of the result.

\subsubsection{High-order and partial mixed field derivatives}

\begin{figure*}[t]
	\centering
		\subfigure[]{\includegraphics[width=5.1cm]{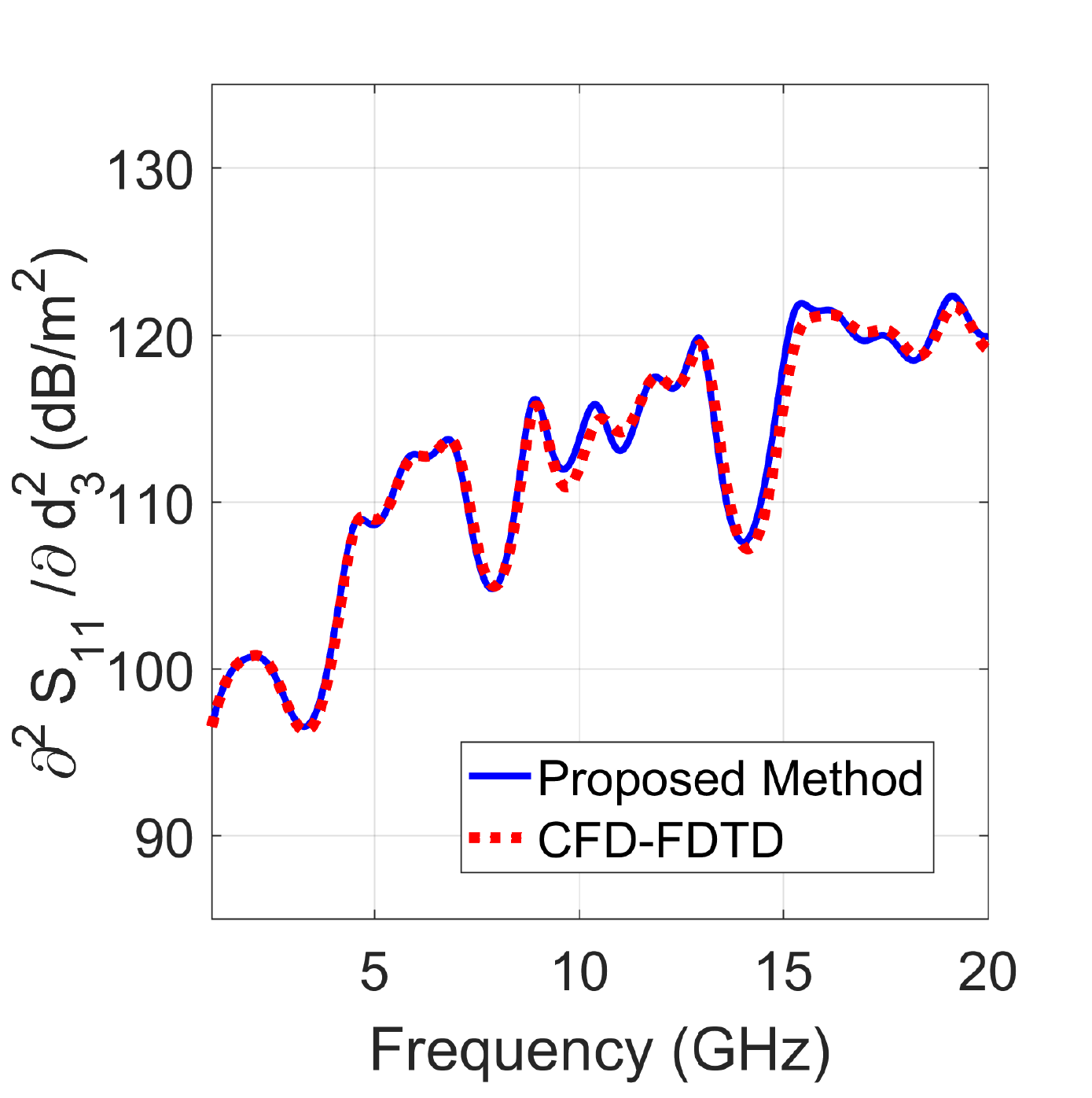}   \label{fig:dfieldL1}}
		\hspace{0.8cm}
		\subfigure[]{\includegraphics[height=5.15cm, width=5.1cm]{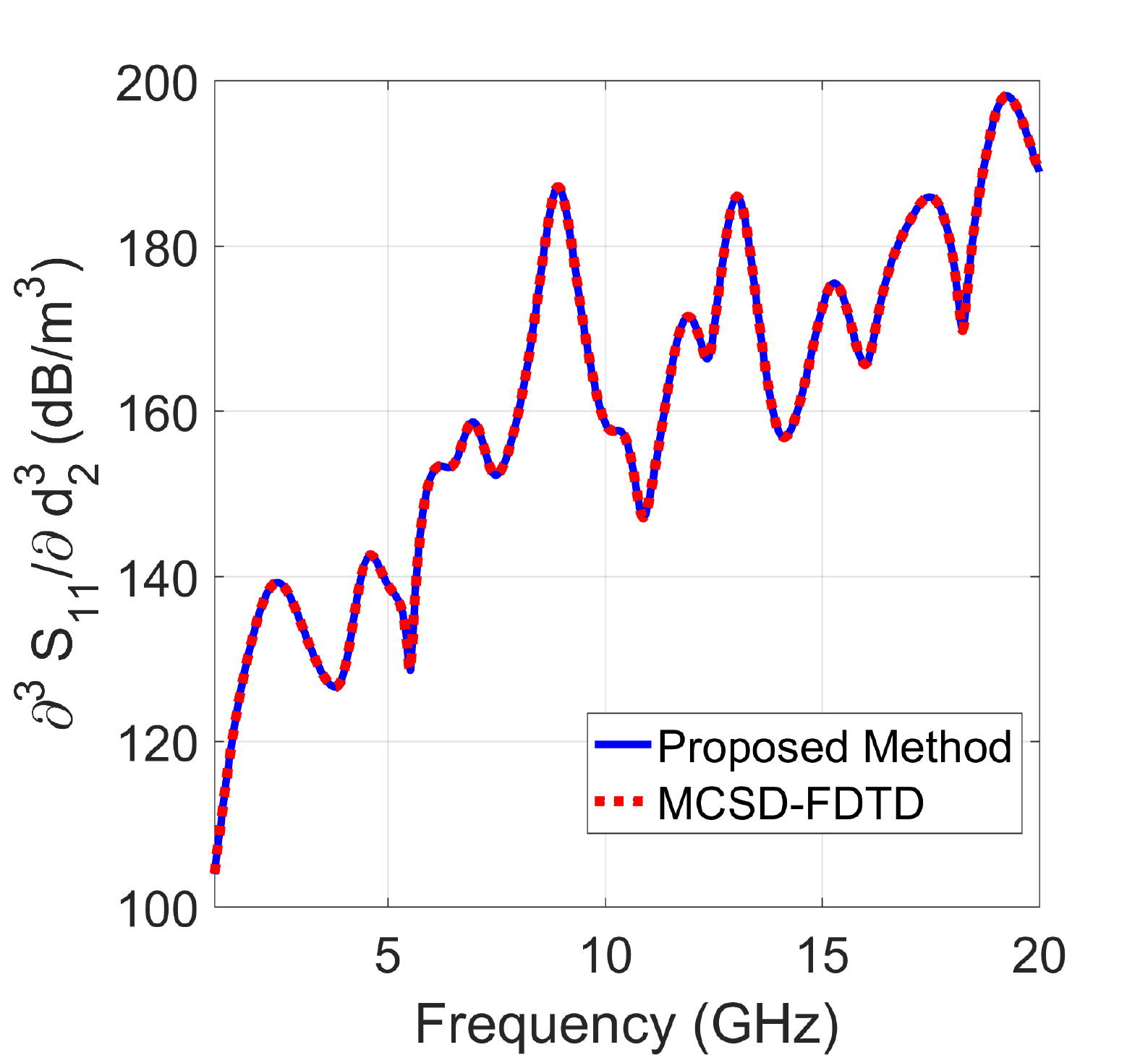} \label{fig:dfieldL1}}
		\hspace{0.8cm}
		\subfigure[]{\includegraphics[width=5.1cm]{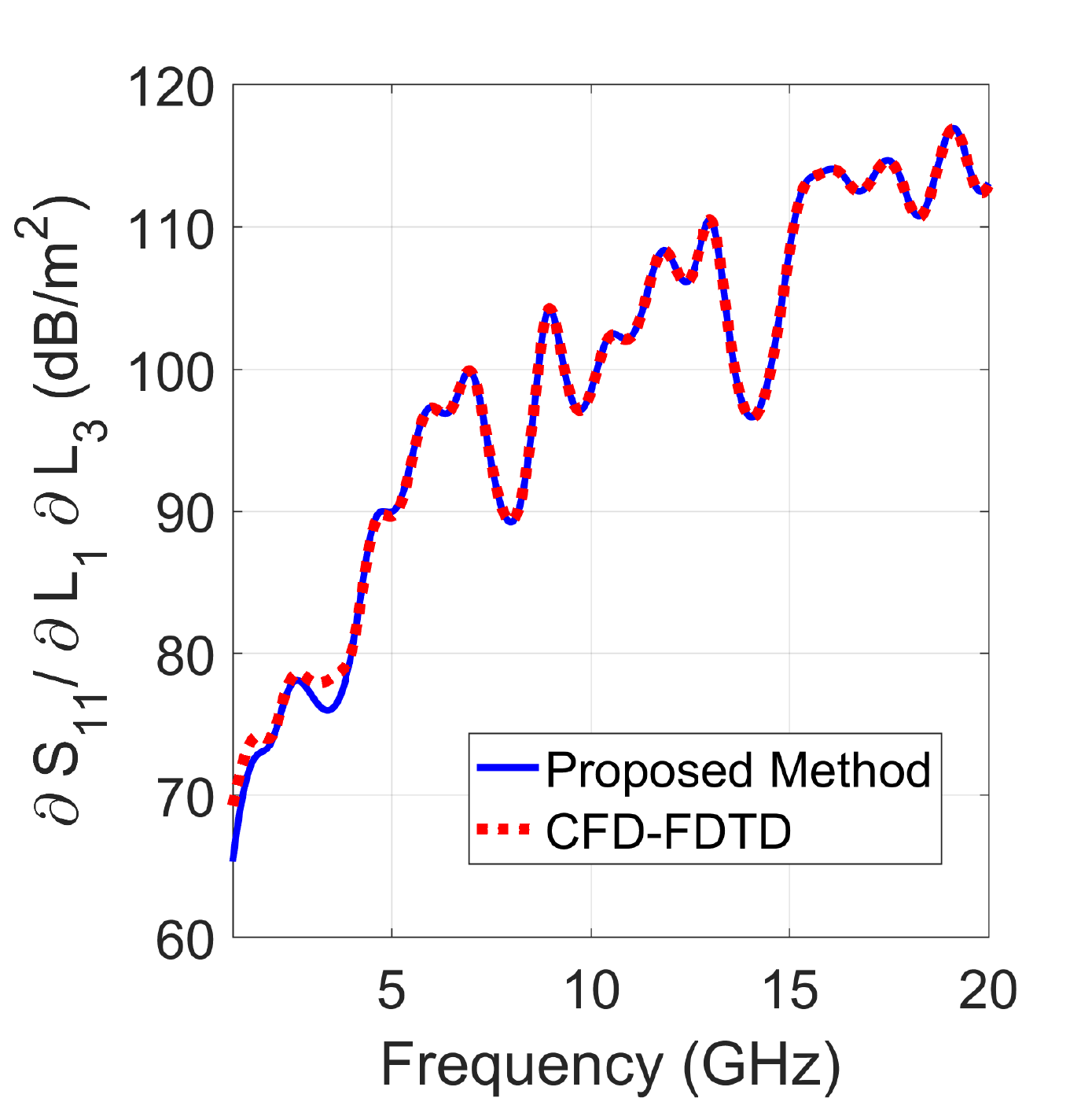} \label{fig:dfieldL1}}		
	\caption{The (a) second-order and (b) third-order derivatives of reflection coefficient $S_{11}$ with respect to $d_3$ and $d_2$ in the frequency domain. In addition, the partial mixed derivatives of $S_{11}$ with respect to both $d_1$ and $d_3$ are shown in (c).}
	\label{ex1_high}
\end{figure*}

The high-order derivatives of the reflected fields are also computed using Algorithm 2. 
For example, the equivalent sources for the second-order field derivatives with respect to $d_3$ are derived by using the transfer function of the cell indexed ${658}$ itself, and the equivalent source for the first-order field derivatives, i.e., $\mathcal{J}_{d_3}(f)$. Fig. \ref{ex1_high} (a) presents the second-order derivatives of $S_{11}$ with respect to $d_3$, computed by using the proposed method and CFD-FDTD. Indeed, higher-order derivatives can be found by recursively using the equivalent sources of lower-order. 

Furthermore, Fig. \ref{ex1_high} (b) shows the third-order derivatives of $S_{11}$ with respect to $d_2$. It should be noted that the subtractive cancellation errors associated with CFD constraint its feasibility in finding third or higher-order derivatives. Even in a 1-D FDTD simulation example, it is difficult to find an appropriate step-size in CFD to compute a convergent result. Alternatively, the Multi-Complex Step Derivative (MCSD)-FDTD method proposed in \cite{mcsd-ims2017} is used to verify the third-order field derivatives in Fig. \ref{ex1_high} (b).  

In addition to high-order field derivatives, the partial mixed field derivatives with respect to any two design parameters are computed using Algorithm 3. Fig. \ref{ex1_high} (c) shows the partial mixed derivative of $S_{11}$ with respect to $d_1$ and $d_3$, in comparison with the CFD-FDTD result. Two full-wave FDTD simulations are needed to derive this second-order derivative using Algorithm 3. The first FDTD is run to compute the transfer function between the cells affected by the perturbations of $d_1$ and of $d_3$. The second FDTD is run sampling the equivalent sources for the field derivatives with respect to $d_1$ and $d_3$. Thus, the equivalent source for this partial mixed field derivatives can be computed by using (\ref{J_cross}). In contrast, at least $4$ FDTD simulations are needed to compute this derivative via CFD-FDTD.
\subsection{Cascaded microstrip filter}
A microstrip filter with three cascaded stubs, shown in Fig.\ref{zoomin} is further studied in this section to demonstrate the proposed methods in three-dimensional FDTD simulations. The structure is discretized with a mesh of $80 \times 150 \times 16$ cells, with $\Delta x = 0.4064$ mm, $\Delta y = 0.4233$ mm,and $\Delta z = 0.265$ mm. The metallic layout is located above a $0.795$ mm thick substrate with relative permittivity $\epsilon_r = 2.2$. The width of microstrips $w = 2.4$ mm and the length of the three stubs $\xi_1 = \xi_2 = \xi_3 = 12.2 $ mm. 

The design parameters of interest in this example are the lengths of each stub: $\xi_1, \xi_2, \xi_3$. A Gaussian pulse $g(t) = \exp (-(t-t_0)^2/T_s^2)$, with $t_0=3 Ts$ and $T_s=15$ ps is used to excite the grid, and $4000$ time steps ($\Delta t = 0.441 $ps) are run. 

\begin{figure}[h]
     \centering
    \includegraphics[width=9.2cm]{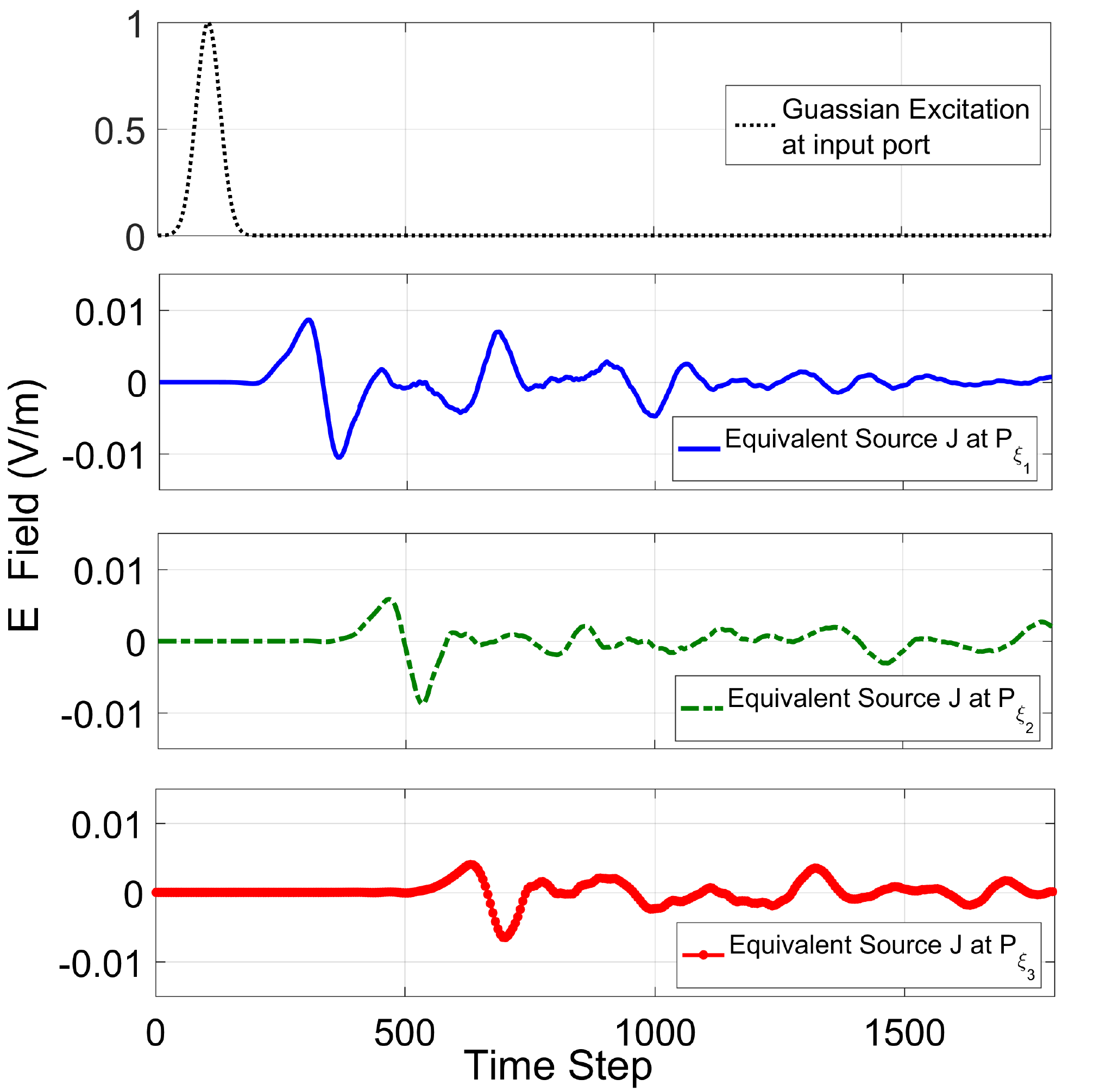}
	\caption{ The Gaussian excitation at the input port of the microstrip filter and the equivalent sources probed at the cells of $P_{\xi_1}, P_{\xi_2}, P_{\xi_3}$ respectively. The order of the first pulse of each equivalent source in time series indicates that the electromagnetic fields reach $P_{\xi_1}:(67,40:45,4), P_{\xi_2}:(67,85:90,4), P_{\xi_3}:(67,120:125,4)$ in sequence.}
\label{ex2_time_domain}
\end{figure}

\subsubsection{First-order field derivatives and time domain analysis}
\begin{figure}[h]
	\centering
	\vspace{-0.24cm}
		\subfigure[]{\includegraphics[width=9cm]{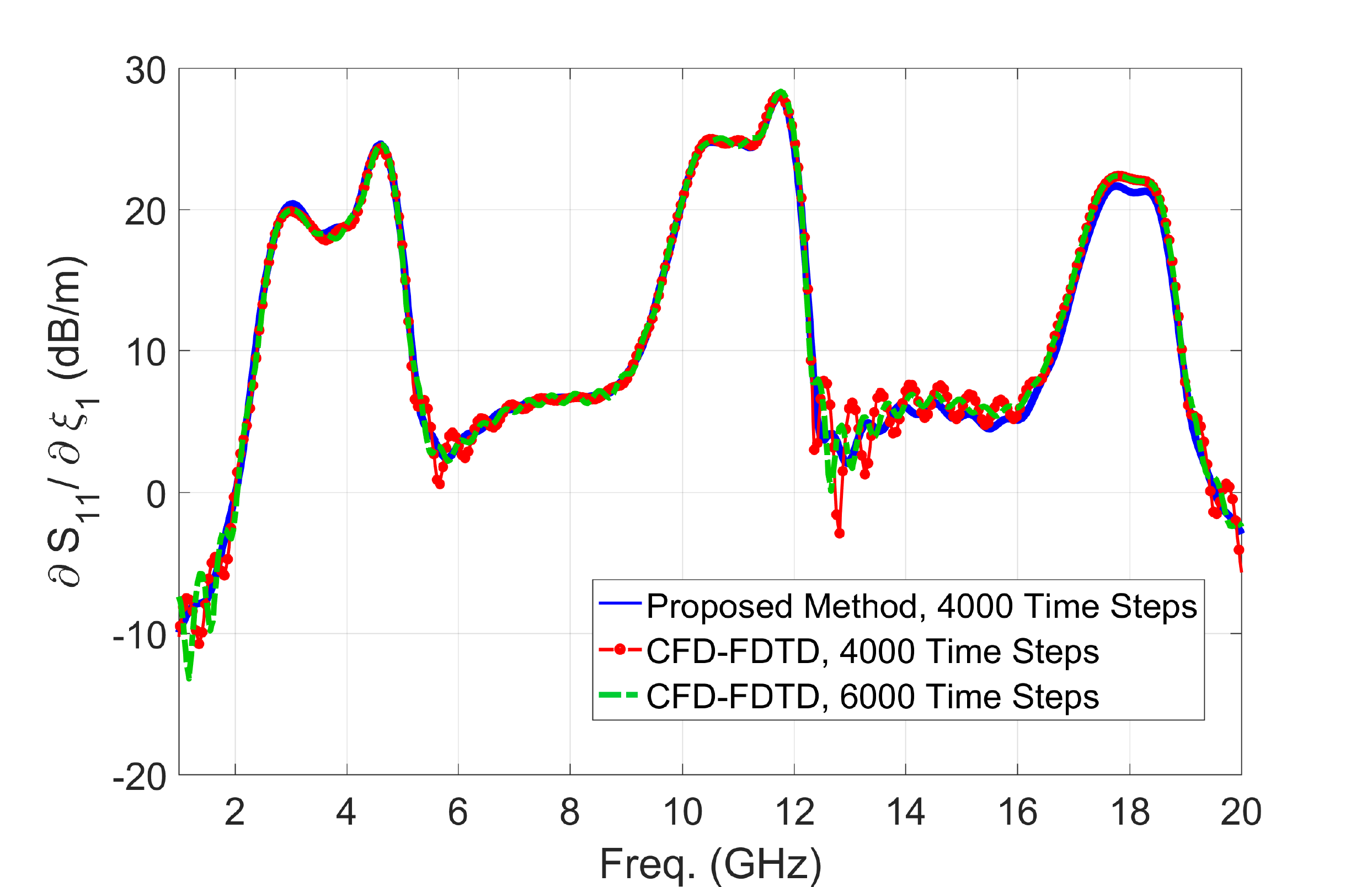}   \label{fig:dfieldL1}}
		\subfigure[]{\includegraphics[width=9cm]{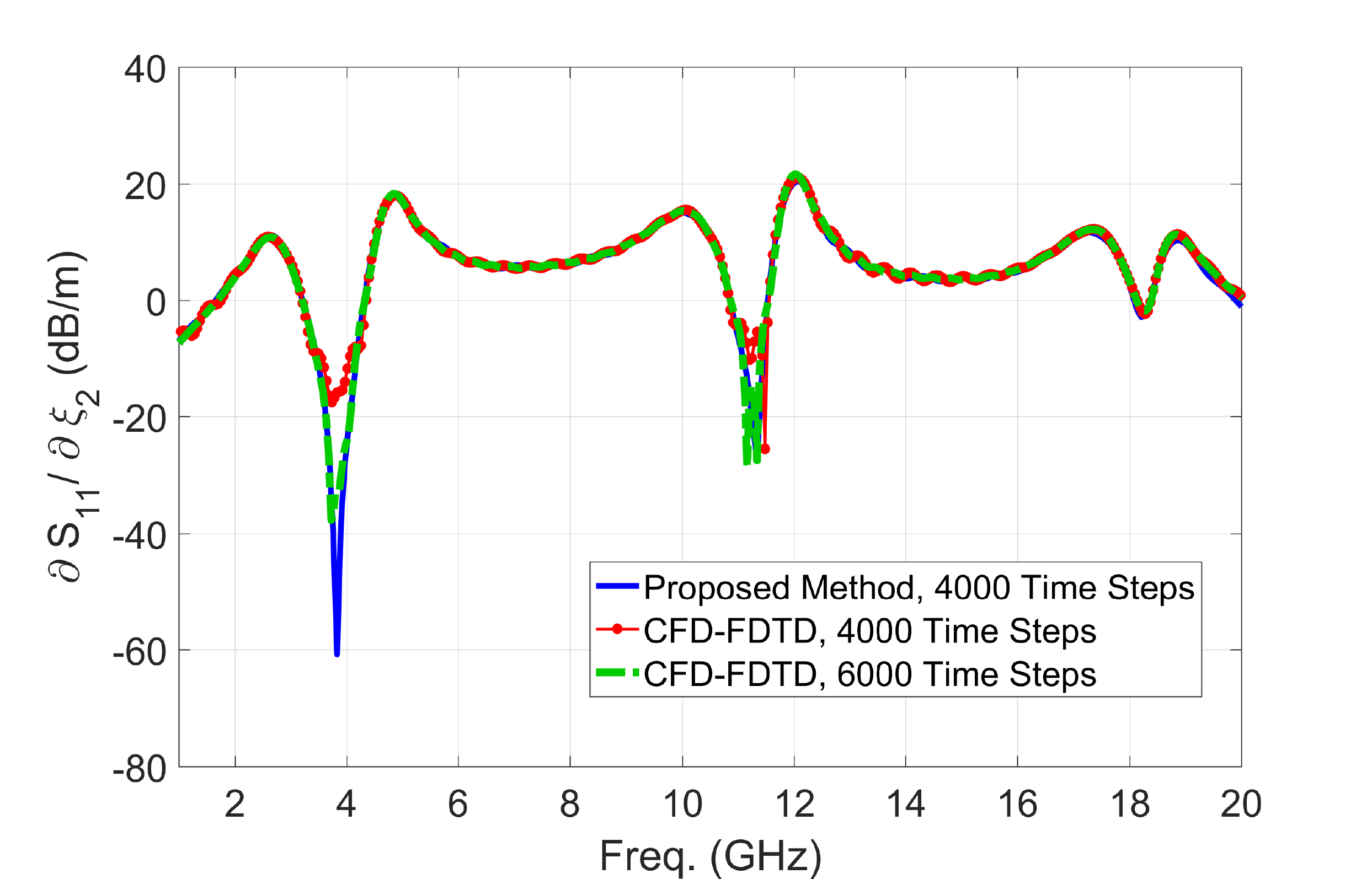} \label{fig:dfieldL1}}		
	\caption{Derivatives of reflection coefficient $S_{11}$, with respect to 
	stub lengths $\xi_1, \xi_2, \xi_3$ in the frequency domain. Results from the proposed method are presented along with those derived with CFD. Significantly, more time steps are needed in CFD-FDTD simulations.}
	\label{ex2_multiparametric}
\end{figure}

In this example, the perturbations of $\xi_1, \xi_2, \xi_3$ are mapped to the $y$-dimension of the cells of subsets $P_{\xi_1}, P_{\xi_2}, P_{\xi_3}$ respectively. Here, $P_{\xi_1}:(67,40:45,4), P_{\xi_2}:(67,85:90,4)$ and $P_{\xi_3}:(67,120:125,4)$. These cells are marked red in Fig. \ref{sparese_matrix}. To find the derivatives of the reflected fields with respect to each parameter, Gaussian excitation is placed at the input port of this filter. Using (\ref{J_DX}), the equivalent sources for field derivatives with respect to $\xi_1, \xi_2, \xi_3$ are computed and recorded at the cells of $P_{\xi_1}, P_{\xi_2}, P_{\xi_3}$. Fig. \ref{ex2_time_domain} shows the time-domain form of this Gaussian excitation and these equivalent sources probed at the cells of  $P_{\xi_1}, P_{\xi_2}, P_{\xi_3}$. By using these equivalent sources and the transfer functions between the input port and the affected cells, the field derivatives at the input port, with respect to $\xi_1, \xi_2, \xi_3$ can be computed. Fig. \ref{ex2_multiparametric} shows the derivatives of the reflection coefficient $S_{11}$ of this filter with respect to $\xi_1$ and $\xi_2$. The results computed by using CFD-FDTD are included to confirm the correctness of this method. Notably, the agreement of the CFD results with those of the proposed method is improved with more time steps. The noise associated in the CFD-FDTD results can be eliminated if the FDTD time-steps are increased from 4000 to 6000.

\begin{figure}[h!]
	\centering
    \includegraphics[width=9cm]{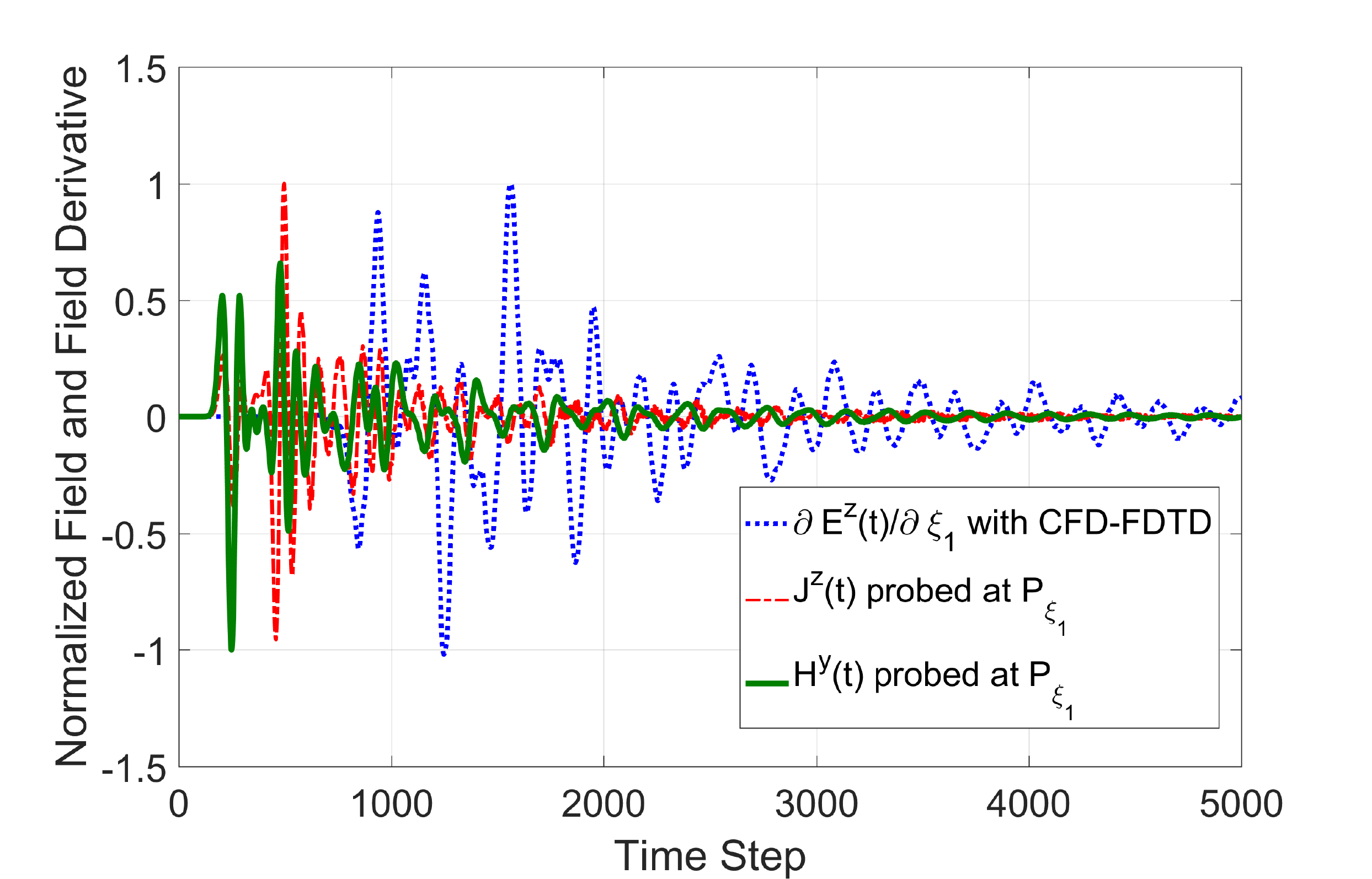}
	\caption{The time domain data of the derivative of electric field with respect to $\xi_1$ computed using CFD-FDTD. Compared to the fields solved directly by FDTD simulation, it takes more time steps for this CFD-FDTD field derivative to reach steady state. On the other hand, the equivalent source computed using the fields directly converges at same order of fields.}
	\vspace{-0.2cm}
	\label{ex2_time_ripple}
\end{figure}

\begin{figure}[h!]
	\centering
	    \subfigure[]{\includegraphics[width=8.5cm]{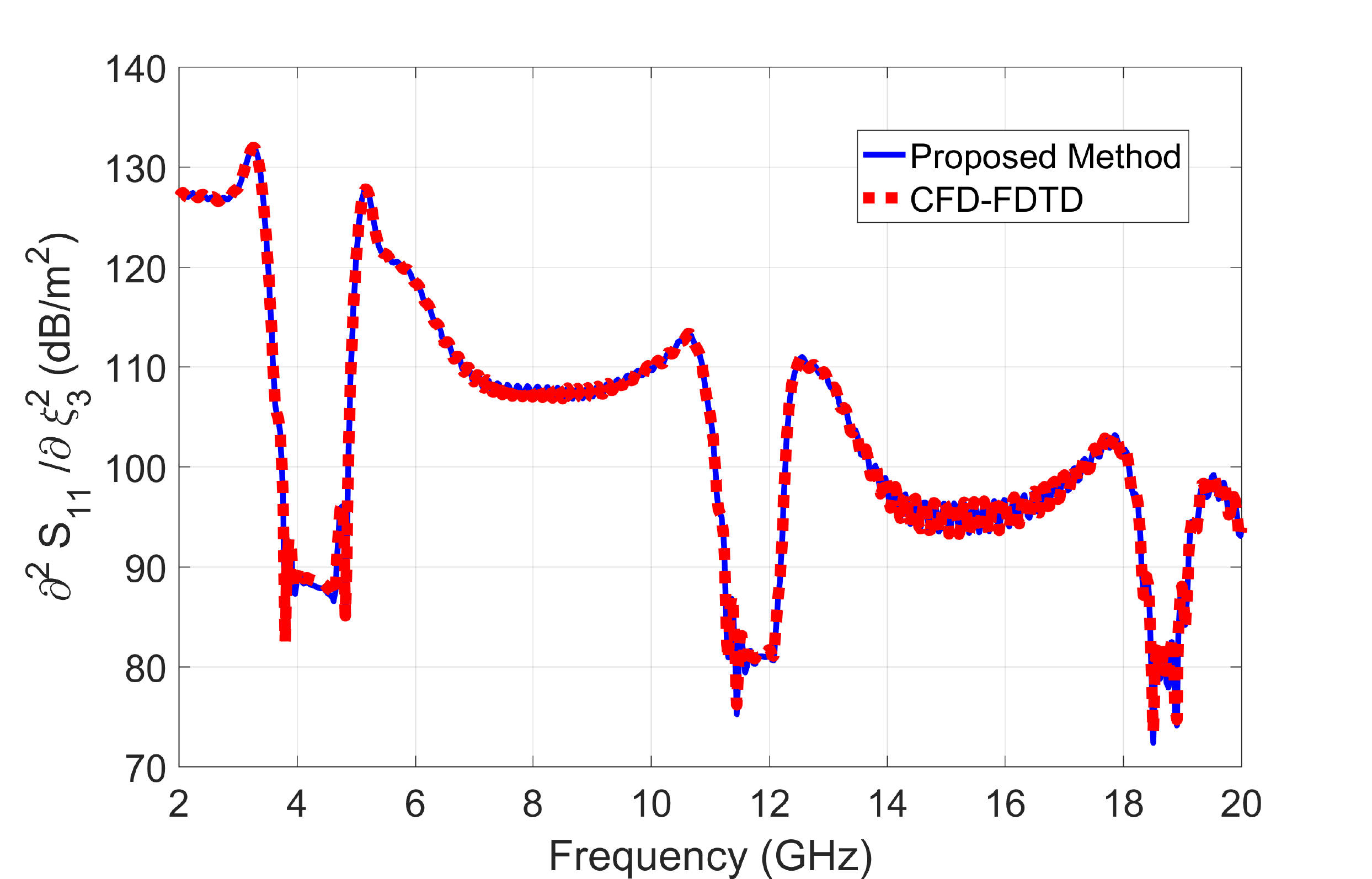} \label{fig:dfieldL1}}
		\subfigure[]{\includegraphics[width=8.5cm]{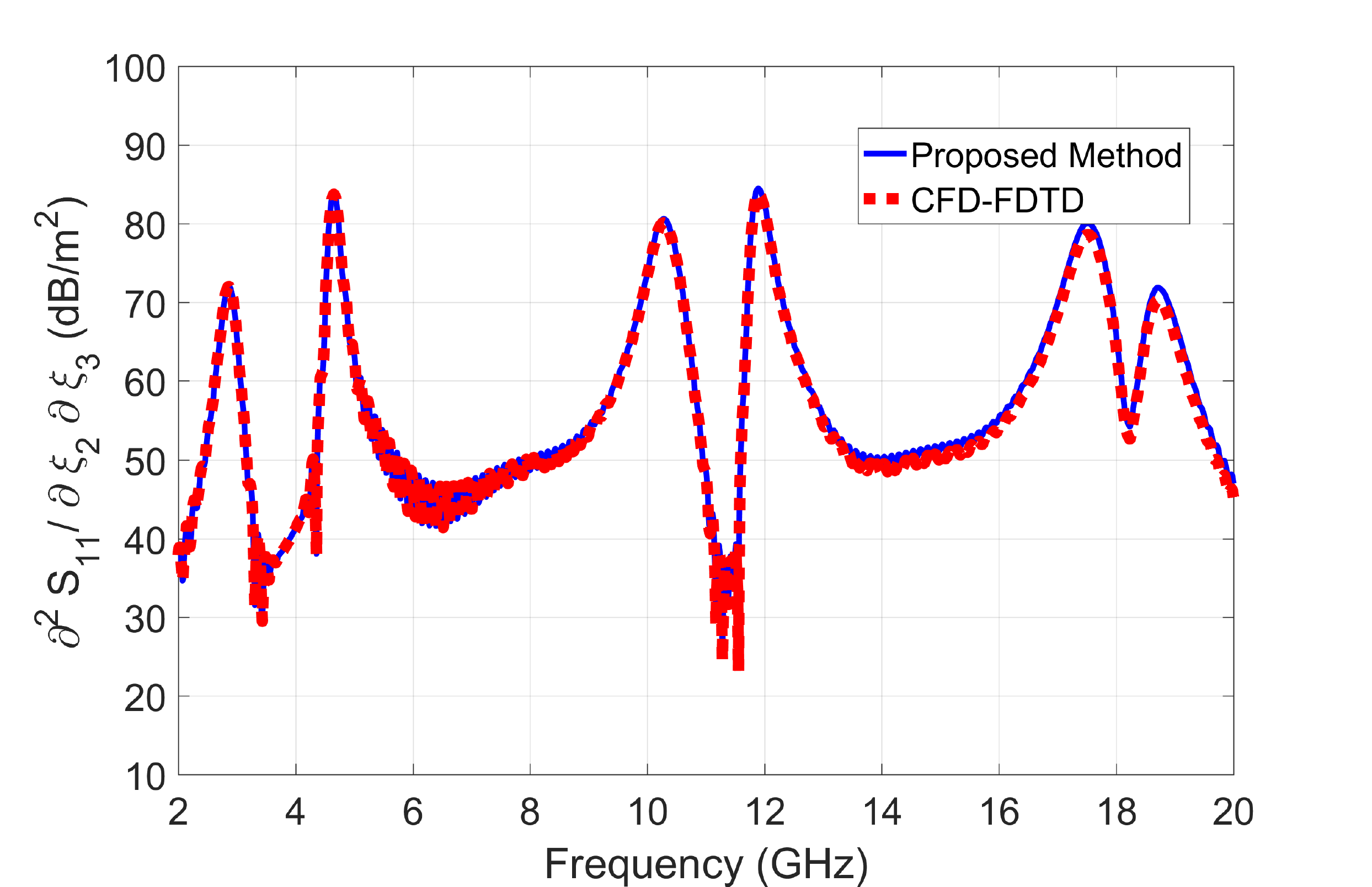}   \label{fig:dfieldL1}}		
	\caption{The (a) second-order derivative of reflection coefficient $S_{11}$, with respect to stub length $\xi_3$ and (b) partial mixed derivative of $S_{11}$ with respect to $\xi_2,\xi_3$ in the frequency domain. CFD-FDTD results are provided in verification.}
	\label{ex2_high}
\end{figure}

The time-domain data of the field derivatives computed by the CFD-FDTD method are  presented in Fig.\ref{ex2_time_ripple} to explain this phenomenon. Compared to the fields determined by one FDTD, it takes more time steps for the field derivatives with CFD-FDTD method to reach steady state. Meanwhile, the equivalent source probed at  $P_{\xi_1}$ is computed by the FDTD fields in each time step. Thus, the equivalent sources converge along with the FDTD field solution and can further produce field derivatives without the noise present in CFD. This indicates that, not only does this method compute field derivatives with guaranteed accuracy, it also requires fewer time steps.

\subsubsection{High-order and partial mixed field derivatives}

Finally, the high-order and partial mixed field derivatives of this filter are computed using Algorithms 2 and 3. For example, Fig. \ref{ex2_high} (a) shows the second-order derivatives of $S_{11}$ with respect to $\xi_3$ compared to the result computed by CFD-FDTD. In addition, Fig. \ref{ex2_high} (b) shows the derivatives of $S_{11}$ with respect to $\xi_2$ and $\xi_3$. These second-order and partial mixed derivatives are the elements of the reflection coefficient Hessian matrix of this structure. Based on the proposed algorithms, a total of 4 FDTD runs are needed to compute the full Hessian matrix, including 3 runs with a Gaussian excitation placed $P_{\xi_1}$, $P_{\xi_2}$, $P_{\xi_3}$ and 1 run with the Gaussian excitation placed at the input port to excite the grid. On the other hand, at least 21 full-wave FDTD runs are needed by using CFD-FDTD. The following table summarizes the number of full-wave FDTD simulation runs with the proposed method, to compute first-order, high-order and partial mixed field derivatives with respect to $N$ design parameters up to $M$-th order.

\begin{table}[h!]
\centering
\caption{Operation Count for The Computation of Field Derivatives in FDTD simulations}
\label{time_table}
\begin{tabular}{|c|c|c|c|c|}
\hline
& $\nabla_{\bs{\xi}}{E^z_{p_a}}$ 
& $\dfrac{\partial^M {E}^z_{p_a}}{\partial \xi^M}$ 
& $\dfrac{\partial^M {E}^z_{p_a}}{\partial \bm{\xi}^M}$ 
& $\mathbb{H}(E^{z}_{p_a})$\\[.5cm] \hline
This paper & $1$ & $2$    & $N+1$    & $N+1$ 
\\\hline
CFD-FDTD & $2N$ &  $(M+1) $ & $N(M+1)$   & $2N^2+N$
\\\hline              
\end{tabular}
\end{table}

\section{Conclusions}
A novel and comprehensive framework to compute field derivatives with respect to multiple design parameters with FDTD up to any order has been presented. The proposed approach utilize standard FDTD simulations, mapping geometrical and material perturbation to equivalent sources. Accuracy-wise, this method is more robust than finite-difference methods, since it is not prone to subtraction errors; moreover, it is more flexible than adjoint variable methods and more scaleable with respect to the number of design variables than complex step methods. Hence, it is a useful addition to the toolbox of time-domain computer-aided analysis and design methods for microwave design. 

\bibliographystyle{IEEEtran}  
\bibliography{mybibfile}  
\appendices

\end{document}